%% file: main.tex
\documentclass{article}

\usepackage{arxiv}

\usepackage[utf8]{inputenc} 
\usepackage[T1]{fontenc}    
\usepackage{hyperref}       
\usepackage{url}            
\usepackage{booktabs}       
\usepackage{amsfonts}       
\usepackage{nicefrac}       
\usepackage{microtype}      
\usepackage[normalem]{ulem}
\usepackage{lineno}

\usepackage{xcolor}
\usepackage{wrapfig}
\usepackage{graphicx,caption}
\usepackage{amsmath,amsfonts,amssymb,amsthm}
\usepackage[toc,page,titletoc]{appendix}
\usepackage{bm}

\newenvironment{keypoints}
{
  \centerline
  {\large \bfseries \scshape Key Points}
  \begin{quote}
}
{
  \end{quote}
}

\newcommand{\beginsupplement}{%
        \setcounter{table}{0}
        \renewcommand{\thetable}{S\arabic{table}}%
        \setcounter{figure}{0}
        \renewcommand{\thefigure}{S\arabic{figure}}%
     }

\DeclareMathOperator*{\argmax}{arg\,max}

\title{Inference of dynamic systems from noisy and sparse data via manifold-constrained Gaussian processes}

\author{
  Shihao Yang \thanks{H. Milton Stewart School of Industrial and Systems Engineering, Georgia Institute of Technology, USA}
   \and
 Samuel W. K. Wong \thanks{Department of Statistics and Actuarial Science, University of Waterloo, Canada}
   \and
   S. C. Kou \thanks{To whom correspondence should be addressed. Department of Statistics, Harvard University, USA. kou@stat.harvard.edu} 
}

\begin{document}

\maketitle

\begin{keypoints}
Ordinary differential equations are a ubiquitous tool for modeling behaviors in science, such as gene regulation, biological rhythms, epidemics and ecology. An important problem is to infer and characterize the uncertainty of parameters that govern the equations. Here we present an accurate and fast inference method using manifold-constrained Gaussian processes, such that the derivatives of the Gaussian process must satisfy the dynamics of the differential equations. Our method completely avoids the use of numerical integration and is thus fast to compute. Our construction is  embedded in a principled statistical framework and is demonstrated to yield fast and reliable inference in a variety of practical problems. Our method works even when some system component(s) is/are unobserved, which is a significant challenge for previous methods.
\end{keypoints}


\newpage

\begin{abstract}
Parameter estimation for nonlinear dynamic system models, represented by ordinary differential equations (ODEs), using noisy and sparse 
data is a vital task in many fields. 
We propose a fast and accurate method, MAGI (MAnifold-constrained Gaussian process Inference), for this task. MAGI uses a Gaussian process model over time-series data, explicitly conditioned on the manifold constraint that derivatives of the Gaussian process must satisfy the ODE system.  By doing so, we completely bypass the need for numerical integration and achieve substantial savings in computational time. MAGI is also suitable for inference with unobserved system components, which often occur in real experiments. MAGI is distinct from existing approaches as we provide a 
principled statistical construction under a Bayesian framework, which incorporates the ODE system through the manifold constraint.  We demonstrate the accuracy and speed of MAGI using realistic examples based on physical experiments.  
\end{abstract}

\section*{Introduction}
Dynamic systems, represented as a set of ordinary differential equations (ODEs), are commonly used to model behaviors in scientific domains, such as gene regulation \cite{hirata2002oscillatory}, \textcolor{black}{biological rhythms \cite{forger2017biological}}, spread of disease \cite{miao2009differential}, ecology \cite{busenberg2012differential}, etc. We focus on models specified by a set of ODEs
\begin{equation}\label{eq:ode}
\dot{\bm{x}}(t) = \frac{d \bm{x}(t)}{dt} = \mathbf{f}(\bm{x}(t),\bm{\theta}, t), \quad t \in [0, T],
\end{equation}
where the vector $\bm{x}(t)$ contains the system outputs that evolve over time $t$, and $\bm{\theta}$ is the vector of model parameters to be estimated from experimental/observational data. 
When $\textbf{f}$ is nonlinear, solving $\bm{x}(t)$ given initial conditions  $\bm{x}(0)$ and $\bm{\theta}$ generally requires a numerical integration method, such as Runge-Kutta.

Historically, ODEs have mainly been used for conceptual or theoretical understanding rather than data fitting as experimental data were limited. Advances in experimental and data-collection techniques have increased the capacity to follow dynamic systems closer to real-time. Such data will generally be recorded at discrete times and subject to measurement error.  
Thus, we assume that we observe $\bm{y}(\bm{\tau}) = \bm{x}(\bm{\tau}) + \bm{\epsilon}(\bm{\tau})$ at a set of observation time points $\bm{\tau}$ 
with error $\bm{\epsilon}$ governed by noise level $\sigma$. Our focus here is inference of $\bm{\theta}$ given $\bm{y}(\bm{\tau})$, with emphasis on nonlinear $\textbf{f}$ where specialized methods that exploit a linear structure, e.g.~\cite{gorbach2017scalable,wu2019parameter}, are not generally applicable.  We shall present a coherent, statistically principled 
framework for dynamic system inference with the help of Gaussian processes (GPs).  The key of our method is to restrict the GPs on a manifold that satisfies the ODE system:  thus we name our method MAGI (MAnifold-constrained Gaussian process Inference).  Placing a GP on $\bm{x}(t)$ facilitates inference of $\bm{\theta}$ without numerical integration, and our explicit manifold constraint is the key novel idea that addresses the conceptual incompatibility between the GP and the specification of the ODE model, as we shall discuss shortly when overviewing our method. We show that the resulting parameter inference is computationally efficient, statistically principled, and effective in a variety of practical scenarios.  MAGI particularly works in the cases when some system component(s) is/are unobserved. To the best of our knowledge, none of the current available software packages \textcolor{black}{that do not use numerical integration} can analyze systems with unobserved component(s).

\subsection*{Overview of our method}

Following the Bayesian paradigm, we view the $D$-dimensional system $\bm{x}(t)$ to be a realization of the stochastic process $\bm{X}(t) = (X_1(t), \ldots, X_D(t))$, and the model parameters $\bm{\theta}$ a realization of the random variable $\bm\Theta$.  In Bayesian statistics, the basis of inference is the posterior distribution, obtained by combining the likelihood function with a chosen prior distribution on the unknown parameters and stochastic processes.  Specifically, we impose a general prior distribution $\pi(\cdot)$ on $\bm{\theta}$ and independent GP prior distributions on each component $X_d(t)$ so that
$ X_d(t) \sim \mathcal{GP}(\mathcal{\mu}_d, \mathcal{K}_d), ~  t \in [0, T]$,
where $\mathcal{K}_d : \mathbb{R}\times\mathbb{R}\to\mathbb{R}$ is a positive definite covariance kernel for the GP and $\mathcal{\mu}_d : \mathbb{R}\to\mathbb{R}$ is the mean function.  Then for any finite set of time points $\bm\tau_d$, $ X_d(\bm\tau_d)$ has a multivariate Gaussian distribution with mean vector $\mathcal{\mu}_d(\bm\tau_d)$ and covariance matrix $\mathcal{K}_d(\bm\tau_d,\bm\tau_d)$.  Denote the observations by $\bm y({\bm\tau}) = (\bm y_1(\bm\tau_1), \ldots, \bm y_D(\bm\tau_D))$, where $\bm \tau = (\bm\tau_1, \bm\tau_2, \dots, \bm\tau_D)$ is the collection of all observation time points and each component $X_d$ can have its own set of observation times $\bm\tau_d = (\tau_{d,1}, \ldots, \tau_{d, N_d})$.
If the $d$-th component is not observed, then $N_d = 0$, and $\bm\tau_d = \emptyset$. $N=N_1 + \cdots +N_D$ is the total number of observations. We note that for the remainder of the paper, the notation $t$ shall refer to time generically, while $\tau$ shall refer specifically to the observation time points.

\begin{figure*}
\centering
\caption{Inference by MAGI for Hes1 partially observed asynchronous system on 2000 simulated datasets. The red curve is the truth.  MAGI recovers the system well, without the usage of any numerical solver:  the green curve shows the median of the inferred trajectories among the 2000 simulated datasets, and a 95\% interval from the 2.5\% and 97.5\% of all inferred trajectories is shown via the blue dashed area. 
}\label{fig:Hes1}

\includegraphics[scale=0.33]{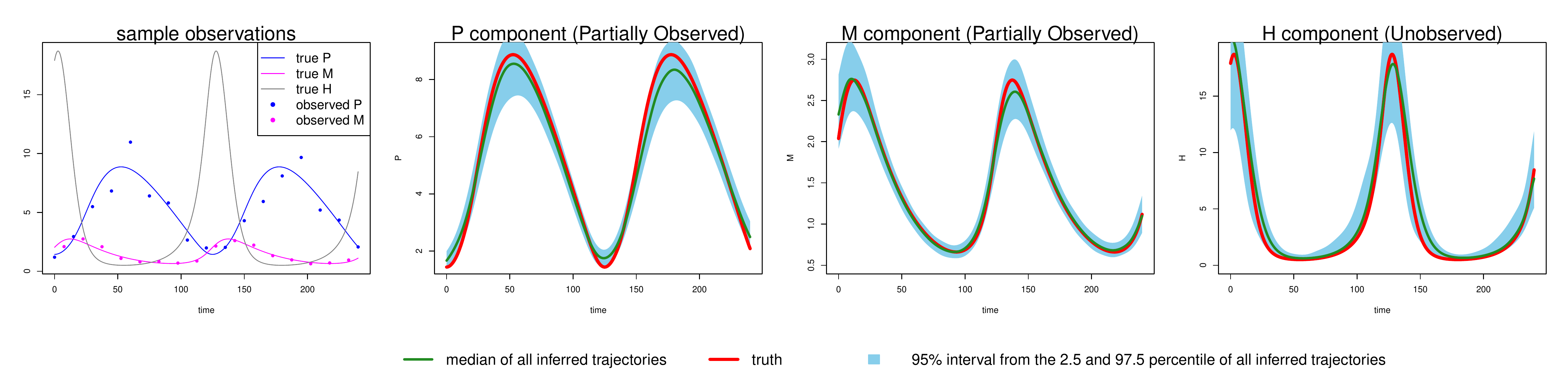}\\
\end{figure*}

As an illustrative example, consider the dynamic system in \cite{hirata2002oscillatory} that governs the oscillation of Hes1 mRNA ($M$) and Hes1 protein ($P$) levels in cultured cells, where it is postulated that a Hes1-interacting ($H$) factor contributes to a stable oscillation, \textcolor{black}{a manifestation of biological rhythm \cite{forger2017biological}}.
The ODEs of the three-component system $X = (P,M,H)$ are
\begin{align*}
\mathbf{f}(X, \bm{\theta}, t) = \begin{pmatrix}
-aPH + bM - cP \\
-dM + \frac{e}{1 + P^2} \\
-aPH + \frac{f}{1+ P^2} - gH
\end{pmatrix},
\end{align*}
where $\bm{\theta} = (a, b, c, d, e, f, g)$ are the associated parameters.
In Fig \ref{fig:Hes1} (left panel) we show noise-contaminated data generated from the system, which closely mimics the experimental setup described in \cite{hirata2002oscillatory}:   $P$ and $M$ are observed at 15-minute intervals for 4 hours but $H$ is never observed. In addition, $P$ and $M$ observations are asynchronous: starting at time 0, every 15 minutes we observe $P$; starting at 7.5 minutes, every 15 minutes we observe $M$; $P$ and $M$ are never observed at the same time. It can be seen that the mRNA and protein levels exhibit the behavior of regulation via negative feedback.
 
The goal here is to infer the seven parameters of the system:  $a, b$ govern the rate of protein synthesis in the presence of the interacting factor; $c,d,g$ are the rates of decomposition; and $e,f$ are inhibition rates. The unobserved $H$ component poses a challenge for most existing \textcolor{black}{methods that do not use numerical integration
}, but is capably handled by MAGI: the $P$ and $M$ panels of Fig \ref{fig:Hes1} show that our inferred trajectories provide good fits to the observed data, and the $H$ panel shows that the dynamics of the entirely unobserved $H$ component are largely recovered as well.  We emphasize that these trajectories are inferred without any use of numerical solvers. 
We shall return to the Hes1 example in detail in the Results section. 

Intuitively, the GP prior on $\bm{X}(t)$ facilitates computation as GP provides closed analytical forms for $\bm{\dot X}(t)$ and $\bm{X}(t)$, which could bypass the need for numerical integration. In particular, with a GP prior on $\bm{X}(t)$, the conditional distribution of $\bm{\dot X}(t)$ given $\bm{X}(t)$ is also a GP with its mean function and covariance kernel completely specified. This GP specification for the derivatives $\bm{\dot x}(t)$, however, is inherently incompatible with the ODE model because \eqref{eq:ode} also completely specifies $\bm{\dot x}(t)$ given $\bm{x}(t)$ (via the function $\mathbf{f}$). As a key novel contribution of our method, MAGI addresses this conceptual incompatibility by constraining the GP to satisfy the ODE model in \eqref{eq:ode}.  To do so, we first define 
a random variable $W$ quantifying the difference between stochastic process $\bm{X}(t)$ and the ODE structure with a given value of the parameter $\bm{\theta}$:
\begin{equation}\label{eq:w}
W = \sup_{t \in [0,T], d \in \{1,\ldots, D\}} |\dot X_d(t) - \mathbf{f}(\bm{X}(t), \bm{\theta}, t)_d|.
\end{equation}
$W \equiv 0$ if and only if ODEs with parameter $\bm{\theta}$ are satisfied by $\bm{X}(t)$. 
Therefore, ideally the posterior distribution for $\bm{X}(t)$ and $\bm{\theta}$ given the observations $\bm y(\bm \tau)$ and \textcolor{black}{the ODE constraint}, $W \equiv 0$, is (informally) 
\begin{equation} \label{eq:ideal-target}
p_{\bm{\Theta}, \bm{X}(t)| W, \bm{Y(\tau}) } (\bm{\theta}, \bm{x}(t) |  W = 0, \bm{Y}(\bm\tau) = \bm{y}(\bm \tau)).
\end{equation}
While \eqref{eq:ideal-target} is the ideal posterior, in reality $W$ is not generally computable. In practice we approximate $W$ by finite discretization on the set $\bm{I} = (t_1, t_2, \ldots, t_n)$ such that $\bm{\tau} \subset \bm{I} \subset [0, T]$ and similarly define $W_{\bm{I}}$ as
\begin{equation}\label{eq:W_I}
W_{\bm{I}} = \max_{t \in \bm{I}, d \in \{1,\ldots, D\}} |\dot X_d(t) - \mathbf{f}(\bm{X}(t), \bm{\theta}, t)_d|.
\end{equation}
Note that $W_{\bm{I}}$ is the maximum of a finite set, and \textcolor{black}{$W_{\bm{I}} \to W$} monotonically as $\bm{I}$ becomes dense in $[0, T]$. Therefore, the practically computable posterior distribution is 
\[
p_{{\bm{\Theta}}, \bm{X}(\bm{I})| W_I, \bm{Y(\tau}) } (\bm{\theta}, \bm{x}(\bm{I}) |  W_{\bm{I}} = 0, \bm{Y}(\bm\tau) = \bm{y}(\bm \tau)),
\]
{\color{black}
which is the joint conditional distribution of $\bm{\theta}$ and $\bm{X}(\bm{I})$ together. Thus, effectively, we simultaneously infer both the parameters and the unobserved trajectory $\bm{X}(\bm{I})$ from the noisy observations $\bm{y}(\bm \tau)$. 

Under Bayes' rule, we have
\begin{align*}
& p_{{\bm{\Theta}}, \bm{X}(\bm{I})| W_I, \bm{Y(\tau}) } (\bm{\theta}, \bm{x}(\bm{I}) |  W_{\bm{I}} = 0, \bm{Y}(\bm\tau) = \bm{y}(\bm \tau)) \\
&\propto  P({\bm{\Theta}} = \bm{\theta}, \bm{X}(\bm{I}) = \bm{x}(\bm{I}), W_{\bm{I}}=0, \bm{Y}(\bm\tau) = \bm{y}(\bm \tau)), \notag\\
\end{align*}
where the right hand side can be decomposed as
{\small
\begin{align*}
& P({\bm{\Theta}} = \bm{\theta}, \bm{X}(\bm{I}) = \bm{x}(\bm{I}), W_{\bm{I}}=0, \bm{Y}(\bm\tau) = \bm{y}(\bm \tau)) \notag\\
&= \pi_{\bm{\Theta}}(\bm{\theta}) \times \underbrace{P(\bm{X}(\bm{I}) = \bm{x({I})}  | {\bm{\Theta}} = \bm{\theta} )}_{(1)} \notag\\
&\qquad \times \underbrace{P(\bm{Y(\bm\tau)} = \bm{y(\bm\tau)} | \bm{X}(\bm I) = \bm{x}(\bm I), {\bm{\Theta}} = \bm{\theta})}_{(2)} \notag\\
&\qquad \times \underbrace{P(W_{\bm{I}}=0 | \bm{Y(\bm\tau)} = \bm{y(\bm\tau)}, \bm{X}(\bm I) = \bm{x}(\bm I), {\bm{\Theta}} = \bm{\theta})}_{(3)}. \notag \\
\end{align*}
} \noindent
The first term (1) can be simplified as
$P(\bm{X}(\bm{I}) = \bm{x({I})}  | {\bm{\Theta}} = \bm{\theta} ) = P(\bm{X}(\bm{I}) = \bm{x({I})} )$
due to the prior independence of $\bm{X}(\bm{I})$ and $\bm\Theta$; it corresponds to the GP prior on $\bm{X}$. The second term (2) corresponds to the noisy observations. 
The third term (3) can be simplified as
{\small 
\begin{align*}
& P(W_{\bm{I}}=0 | \bm{Y(\bm\tau)} = \bm{y(\bm\tau)}, \bm{X}(\bm I) = \bm{x}(\bm I), {\bm{\Theta}} = \bm{\theta}) \\
& =P(\bm{\dot X({I})} - \mathbf{f}(\bm{x}(\bm{I}), \bm{\theta}, t_{\bm{I}}) = \bm 0 | \bm{Y(\bm\tau)} = \bm{y(\bm\tau)}, \bm{X}(\bm I) = \bm{x}(\bm I), {\bm{\Theta}} = \bm{\theta}) \\
&= P(\bm{\dot X({I})} - \mathbf{f}(\bm{x}(\bm{I}), \bm{\theta}, t_{\bm{I}}) = \bm 0 | \bm{X}(\bm{I}) = \bm{x({I})}) \\
&= P(\bm{\dot X({I})} = \mathbf{f}(\bm{x}(\bm{I}), \bm{\theta}, t_{\bm{I}}) | \bm{X}(\bm{I}) = \bm{x({I})}),
\end{align*}
}
which is the conditional density of $\bm{\dot X({I})}$ given $\bm{X}(\bm{I})$ evaluated at $\mathbf{f}(\bm{x}(\bm{I}), \bm{\theta}, t_{\bm{I}})$. All three terms are multivariate Gaussian: the third term is Gaussian 
because $\bm{\dot X({I})}$ given $\bm{X}(\bm{I})$ has a multivariate Gaussian distribution as long as the kernel $\mathcal{K}$ is twice differentiable.
} 

\textcolor{black}{Therefore,} 
the practically computable posterior distribution simplifies to 
{\small 
\begin{align}
& p_{{\bm{\Theta}}, \bm{X}(\bm{I})| W_I, \bm{Y(\tau}) } (\bm{\theta}, \bm{x}(\bm{I}) |  W_{\bm{I}} = 0, \bm{Y}(\bm\tau) = \bm{y}(\bm \tau)) \label{eq:real-target} \\
&\propto \pi_{\bm{\Theta}}(\bm{\theta}) \exp\Big\{-\frac{1}{2}\sum_{d=1}^D \Big[ \notag\\
& + \underbrace{ |\bm{I}|\log(2\pi) + \log{|C_d|} + \left\|x_d(\bm{I}) - \mu_d(\bm{I})\right\|_{C_d^{-1}}^2}_{(1)} \notag\\
& + \underbrace{ |\bm{I}|\log(2\pi) + \log{|K_d|} + \left\|\mathbf{f}_{d, \bm{I}}^{\bm{x}, \bm{\theta}} - \dot{\mu}_d(\bm{I}) - m_d \{x_d(\bm{I}) - \mu_d(\bm{I})\}\right\|_{K_d^{-1}}^2}_{(3)}\notag \\
& + \underbrace{ N_d \log(2\pi \sigma_d^{2}) +  \left\|x_d(\bm{\tau}_d) - y_d(\bm{\tau}_d)\right\|_{\sigma_d^{-2} }^2 }_{(2)} \Big]\Big\}  \notag 
\end{align}
}%
where $\|\bm v\|_A^2 = \bm{v}^\intercal A \bm{v}$, $|\bm{I}|$ is the cardinality of $\bm{I}$, $\mathbf{f}_{d, \bm{I}}^{\bm{x}, \bm{\theta}}$ is short for the $d$-th component of $\mathbf{f}(\bm{x}(\bm{I}), \bm{\theta}, t_{\bm{I}})$, and the multivariate Gaussian covariance matrix $C_d$ and the matrix $K_d$ can be derived as follows for each component $d$:
\begin{equation}
\begin{cases}
C &= \mathcal{K}(\bm{{I}}, \bm{I}) \\
m &= \mathcal{'K}(\bm{I}, \bm{I}) \mathcal{K}(\bm{I}, \bm{I})^{-1} \\
K &= \mathcal{K''}(\bm{I}, \bm{I}) - \mathcal{'K}(\bm{I}, \bm{I}) \mathcal{K}(\bm{I}, \bm{I})^{-1} \mathcal{K'}(\bm{I}, \bm{I})
\end{cases}
\end{equation}
where $\mathcal{'K} = \frac{\partial}{\partial s} \mathcal{K}(s, t)$, $\mathcal{K'} = \frac{\partial}{\partial t} \mathcal{K}(s, t)$, and $\mathcal{K''} = \frac{\partial^2}{\partial s\partial t} \mathcal{K}(s, t)$.

In practice we choose the Matern kernel $\mathcal{K}(s, t) = \phi_1\frac{2^{1-\nu}}{\Gamma(\nu)}\left(\sqrt{2\nu}\frac{l}{\phi_2}\right)^\nu B_\nu\left(\sqrt{2\nu}\frac{l}{\phi_2}\right)$ where $l = |s-t|$, $\Gamma$ is the Gamma function and $B_\nu$ is the modified Bessel function of the second kind, and the degree of freedom $\nu$ is set to be 2.01 to ensure that the kernel is twice differentiable. $\mathcal{K}$ has two hyper-parameters $\phi_1$ and $\phi_2$. Their meaning and specification are discussed in the \textit{Materials and Methods} section.

{\color{black}
With the posterior distribution specified in \eqref{eq:real-target}, we use Hamiltonian Monte Carlo (HMC) \cite{neal2011mcmc} to obtain samples of $\bm{X_I}$ and the parameters together.  
At the completion of HMC sampling, we take the posterior mean of $\bm{X_I}$ as the inferred trajectory, and the posterior means of the sampled parameters as the parameter estimates. Throughout the MAGI computation, no numerical integration is ever needed. 
} 

\subsection*{Review of related work}
The problem of dynamic system inference has been studied in the literature, which we now briefly review.  We first note that a simple approach to constructing the `ideal' likelihood function is according to $p(\bm{y}(\bm{t}) | \hat{\bm{x}}(\bm{t},\bm{\theta}, \bm{x}(0) ), \sigma)$, where $\hat{\bm{x}}(\bm{t},\bm{\theta},\bm{x}(0))$ is the numerical solution of the ODE obtained by numerical integration given $\bm{\theta}$ and the initial conditions. This approach suffers from a high computational burden: numerical integration is required for every $\bm{\theta}$ sampled in an optimization or Markov chain Monte Carlo (MCMC) routine \cite{calderhead2009accelerating}. Smoothing methods have been useful for eliminating the dependence on numerical ODE solutions, and an innovative penalized likelihood approach \cite{ramsay2007parameter} uses a B-spline basis for constructing estimated functions to simultaneously satisfy the ODE system and fit the observed data.  While in principle the method in \cite{ramsay2007parameter} can handle an unobserved system component, substantive manual input is required as we show in the Results, which contrasts with the ready-made solution that MAGI provides.

As an alternative to the penalized likelihood approach, GPs are a natural candidate for fulfilling the smoothing role in a Bayesian paradigm due to their flexibility and analytic tractability \cite{hennig2015probabilistic}.  The use of GPs to approximate the dynamic system and facilitate computation has been previously studied by a number of authors \cite{calderhead2009accelerating,dondelinger2013ode,barber2014gaussian,ghosh2017fast,lazarus2018multiphase,pmlr-v89-wenk19a}.
The basic idea is to specify a joint GP over $\bm{y}, \bm{x}, \dot{\bm{x}}$ with hyperparameters $\phi$, and then provide a factorization of the joint density $p(\bm{y}, \bm{x}, \dot{\bm{x}},\bm{\theta},\phi,\sigma)$ that is suitable for inference.  The main challenge is to find a coherent way to combine information from two distinct sources: the approximation to the system by the GP governed by hyperparameters $\phi$, and the actual dynamic system equations governed by parameters $\bm{\theta}$. In \cite{calderhead2009accelerating,dondelinger2013ode}, the factorization proposed is $p(\bm{y}, \bm{x}, \dot{\bm{x}},\bm{\theta},\phi,\sigma) = p( \bm{y} | \bm{x}, \sigma) p(\dot{\bm{x}} | \bm{x}, \bm{\theta}, \phi)  p(\bm{x} | \phi)  p( \phi ) p( \bm{\theta} )$, where $p( \bm{y} | \bm{x}, \sigma)$ comes from the observation model and  $p(\bm{x} | \phi)$ comes from the GP prior as in our approach.  However, there are significant conceptual difficulties in specifying $p(\dot{\bm{x}} | \bm{x}, \bm{\theta}, \phi)$: on one hand, the distribution of $\dot{\bm{x}}$ is completely determined by the GP given $\bm{x}$, 
while on the other hand $\dot{\bm{x}}$ is completely specified by the ODE system $\dot{\bm{x}} = \mathbf{f}(\bm{x}, \bm{\theta}, t)$; these two are incompatible.
Previous authors have attempted to circumvent this incompatibility of the GP and ODE system: 
\cite{calderhead2009accelerating,dondelinger2013ode} use a product-of-experts heuristic by letting $p(\dot{\bm{x}} | \bm{x}, \bm{\theta}, \phi) \propto p(\dot{\bm{x}} | \bm{x}, \phi) p(\dot{\bm{x}} | \bm{x}, \bm{\theta})$, where the two distributions in the product come from the GP and a noisy version of the ODE, respectively. In \cite{pmlr-v89-wenk19a}, the authors arrive at the same posterior as \cite{calderhead2009accelerating,dondelinger2013ode} by assuming an alternative graphical model that bypasses the product of experts heuristic; nonetheless, the method requires working with an artificial noisy version of the ODE.   In \cite{barber2014gaussian}, the authors start with a different factorization:  $p(\bm{y}, \bm{x}, \dot{\bm{x}},\bm{\theta},\phi,\sigma)=p( \bm{y} | \dot{\bm{x}}, \phi,\sigma) p(\dot{\bm{x}} | \bm{x}, \bm{\theta}) p(\bm{x} | \phi)p( \phi ) p( \bm{\theta} )$, where $p( \bm{y} | \dot{\bm{x}}, \phi)$ and $p(\bm{x} | \phi)$ are given by the GP and $p(\dot{\bm{x}} | \bm{x}, \bm{\theta})$  is a Dirac delta distribution given by the ODE.  However, this factorization is incompatible with the observation model $p(\bm{y}| \bm{x},\sigma)$ as discussed in detail in \cite{macdonald2015controversy}.  There is other related work that uses GPs in an ad hoc partial fashion to aid inference.  In \cite{ghosh2017fast}, GP regression is used to obtain the means of $\bm{x}$ and $\dot{\bm{x}}$ for embedding within an Approximate Bayesian Computation estimation procedure. In \cite{lazarus2018multiphase}, GP smoothing is used during an initial burn-in phase as a proxy for the likelihood, before switching to the `ideal' likelihood to obtain final MCMC samples. While empirical results from the aforementioned 
studies are promising, a principled statistical framework for inference that addresses the previously noted conceptual incompatibility between the GP and ODE specifications is lacking.  
Our work presents one such principled statistical framework through the explicit manifold constraint.  MAGI is therefore distinct from recent GP-based approaches \cite{dondelinger2013ode,pmlr-v89-wenk19a} or any other Bayesian analogs of \cite{ramsay2007parameter}.

In addition to the conceptual incompatibility, none of the existing  methods \textcolor{black}{that do not use numerical integration} offer a practical 
solution for a system with unobserved component(s), which highlights another unique and important contribution of our approach. 
%

\section*{Results}
We apply MAGI to three systems. 
We begin with an illustration that demonstrates the effectiveness of MAGI in practical problems with unobserved system component(s). Then, we make comparisons with other current methods on two benchmark systems, which show that our proposed method provides more accurate inference while having much faster runtime.


\subsection*{Illustration: Hes1 model}

The Hes1 model described in the Introduction demonstrates inference on a system with an unobserved component and asynchronous observation times.
This section continues the inference of this model. Ref \cite{hirata2002oscillatory} studied the theoretical oscillation behavior using parameter values $a = 0.022$, $b = 0.3$, $c = 0.031$, $d = 0.028$; $e = 0.5$, $f = 20$,  $g = 0.3$, which leads to one oscillation cycle approximately every 2 hours. Ref \cite{hirata2002oscillatory} also set the initial condition at the lowest value of $P$ when the system is in oscillation equilibrium \cite{hirata2002oscillatory}: $P = 1.439$, $M = 2.037$, $H = 17.904$. The noise level in our simulation is derived from \cite{hirata2002oscillatory} where the standard error based on repeated measures are reported to be around 15\% of the $P$ (protein) level and $M$ (mRNA) level, so we set the simulation noise to be multiplicative following a log-normal distribution with standard deviation 0.15, and throughout this example we assume the noise level $\sigma$ is known to be 0.15 from repeated measures reported in \cite{hirata2002oscillatory}. The $H$ component is never observed. Owing to the multiplicative error on the strictly positive system, we apply our method to the log-transformed ODEs, so that the resulting error distributions are Gaussian. To the best of our knowledge, MAGI is the only one that provides a practical and complete solution for handling \textcolor{black}{unobserved} component cases like this example.




We generate 2000 simulated datasets based on the above setup for the Hes1 system. The left-most panel in Fig \ref{fig:Hes1} shows one example dataset. For each dataset, we use MAGI to infer the trajectories and estimate the parameters.  
We use the posterior mean of $X_t = (P, M, H)_t$ as the inferred trajectories for the system components, which are generated by MAGI without using any numerical solver. Fig \ref{fig:Hes1} summarizes the inferred trajectories across the 2000 simulated datasets, showing the median of the inferred trajectories of $X_t$ together with the 95\% interval of the inferred trajectories represented by the 2.5\% and 97.5\% percentiles. 
The posterior mean of $\bm{\theta} = (a,b,c,d,f,e,g)$ is our estimate of the parameters. Table \ref{tab:Hes1} summarizes the parameter estimates across the 2000 simulated datasets, by showing their means and standard deviations. Fig \ref{fig:Hes1} shows that MAGI recovers the system well, including the completely unobserved $H$ component. 
Table \ref{tab:Hes1} shows that MAGI also recovers the system parameters well, except for the parameters that only appear in the equation for the unobserved $H$ component, which we will discuss shortly.  Together, Fig \ref{fig:Hes1} and Table \ref{tab:Hes1} demonstrate that MAGI can recover the entire system without any usage of a numerical solver, even in the presence of unobserved component(s). 

\subsubsection*{Metrics for assessing the quality of  system recovery} 
To further assess the quality of the parameter estimates and the system recovery, we consider two metrics.  First, as shown in Table \ref{tab:Hes1}, we examine the accuracy of the parameter estimates by directly calculating the  root mean squared error (RMSE) of the parameter estimates to the true parameter value. We call this measure the \emph{parameter RMSE} metric. Second, it is possible that a system might be insensitive to some of the parameters; in the extreme case, some parameters may not be fully identifiable given only the observed data and components. In these situations, it is possible that the 
system trajectories implied by quite distinct parameter values are similar to each other (or even close to the true trajectory).  We thus consider an additional \textit{trajectory RMSE} metric to account for possible parameter insensitivity, and measure how well the system components are recovered given the parameter and initial condition estimates.
The trajectory RMSE is obtained by treating the numerical ODE solution based on the true parameter value as the ground truth:  first, the numerical solver is used to reconstruct the trajectory based on the estimates of the parameter and initial condition (from a given method); then, we calculate the RMSE of this reconstructed trajectory to the true trajectory at the observation time points.
We emphasize that the trajectory RMSE metric is only for evaluation purpose to assess (and compare across methods) how well a method recovers the trajectories of the system components, and that throughout MAGI no numerical solver is ever needed.


We summarize the trajectory RMSEs of MAGI in Table \ref{tab:Hes1-rmse} for the Hes1 system.

We compare MAGI with the benchmark provided by the
B-spline-based penalization approach of Ref
\cite{ramsay2007parameter}. To the best of our knowledge, \textcolor{black}{among all the existing methods that do not use numerical integration}, Ref \cite{ramsay2007parameter} is the only one with a software package that can be manually adapted to handle an unobserved component.  We note, however, this package itself is not 
ready-made for this problem: it requires substantial manual input as it does not have default or built-in setup of its hyper-parameters for the unobserved component. 
None of the other benchmark methods, \textcolor{black}{including Ref \cite{dondelinger2013ode,pmlr-v89-wenk19a}}, 
provide software that is equipped to handle an unobserved component.
Table \ref{tab:Hes1} compares our estimates against those given by Ref
\cite{ramsay2007parameter} based on the parameter RMSE, which shows that the parameter RMSEs for MAGI are substantially smaller than
\cite{ramsay2007parameter}.
Fig \ref{fig:Hes1} shows that the inferred trajectories from MAGI are very close to the truth. 
On the contrary, the method in \cite{ramsay2007parameter} is not able to recover the \textcolor{black}{unobserved} component $H$ nor the associated parameter $f$ and $g$; see Fig S1 in the SI for the plots. Table \ref{tab:Hes1-rmse} compares the trajectory RMSE of the two methods. It is seen that the trajectory RMSE of MAGI is substantially smaller than that of \cite{ramsay2007parameter}. Further implementation details and comparison 
are provided in the SI. 


Finally, we note that MAGI recovers the unobserved component $H$ almost as well as the observed components of $P$ and $M$, as measured by the trajectory RMSEs. In comparison, for the result of \cite{ramsay2007parameter} in Table \ref{tab:Hes1-rmse}, the trajectory RMSE of the unobserved $H$ component is orders of magnitude worse than those of $P$ and $M$. The numerical results thus illustrate the effectiveness of MAGI in borrowing information from the observed components to infer the unobserved component, which is made possible by explicitly conditioning on the ODE structure. The self-regulating parameter $g$ and inhibition rate parameter $f$ for the unobserved component appear to have high inference variation across the simulated datasets despite the small trajectory RMSEs. 
This suggests that the system itself could be insensitive to $f$ and $g$ 
when the $H$ component is unobserved.



\begin{table}[ht]
\centering
\caption{Parameter inference in the Hes1 partially observed asynchronous system based on 2000 simulation datasets. 
Average parameter estimates based on MAGI and Ref \cite{ramsay2007parameter} across the 2000 simulated datasets are reported together with the standard deviation. Parameter RMSEs are reported in the following column. The boldface highlights the best method in terms of parameter RMSE for each parameter. 
}\label{tab:Hes1}

\begin{tabular}{cc|rr|rr}
  \hline
 & & \multicolumn{2}{c|}{MAGI} & \multicolumn{2}{c}{Ref \cite{ramsay2007parameter}} \\
 $\bm{\theta}$ & Truth & Estimate & RMSE & Estimate & RMSE \\
  \hline
  a & 0.022 &  0.021 $\pm$ 0.003   & \textbf{0.003} & 0.027 $\pm$ 0.026 & 0.026 \\ 
  b & 0.3 &    0.329 $\pm$ 0.051   & \textbf{0.059} & 0.302 $\pm$ 0.086 & 0.086 \\ 
  c & 0.031 &  0.035 $\pm$ 0.006   & \textbf{0.007} & 0.031 $\pm$ 0.010 & 0.010 \\ 
  d & 0.028 &  0.029 $\pm$ 0.002   & \textbf{0.003} & 0.028 $\pm$ 0.003 & \textbf{0.003} \\ 
  e & 0.5 &    0.552 $\pm$ 0.074   & 0.090 & 0.498 $\pm$ 0.088 & \textbf{0.088} \\ 
  f & 20 &    13.759 $\pm$ 3.026   & \textbf{6.936} & 604.9 $\pm$ 5084.8 & 5117.0 \\ 
  g & 0.3 &    0.141 $\pm$ 0.026   & \textbf{0.162} & 1.442 $\pm$ 9.452 & 9.519 \\   
  \hline
\end{tabular}

\end{table}

\begin{table}[ht]
\centering
\caption{Trajectory RMSEs of the individual components in the Hes1 system, comparing the average trajectory RMSEs of MAGI and Ref \cite{ramsay2007parameter} over the 2000 simulated datasets.  The best trajectory RMSE for each system component is shown in boldface.
}\label{tab:Hes1-rmse}

\begin{tabular}{llll}
  \hline
Method & $P$ & $M$ & $H$\\
  \hline
MAGI & \textbf{0.97} & \textbf{0.21} & \textbf{2.57 }\\
Ref \cite{ramsay2007parameter} & 1.30 & 0.40 & 59.47 \\
   \hline
\end{tabular}
\end{table}





\subsection*{Comparison with previous methods based on GPs}
To further assess MAGI, we compare with two methods: Adaptive Gradient Matching (AGM) of Ref \cite{dondelinger2013ode} and Fast Gaussian process based Gradient Matching (FGPGM) of Ref \cite{pmlr-v89-wenk19a}, representing the state-of-the-art of inference methods based on GPs. 
For fair comparison, we use the same benchmark systems, scripts and software provided by the authors for performance assessment, and run the software using the settings recommended by the authors. The benchmark systems include the FitzHugh-Nagumo (FN) equations \cite{fitzhugh1961impulses} and a protein transduction model \cite{vyshemirsky2007bayesian}. 


\subsubsection*{FN model}
The FitzHugh-Nagumo (FN) equations are a classic Ion channel model that describes spike potentials. The system consists of $X = (V,R)$, where $V$ is the variable defining the voltage of the neuron membrane potential and $R$ is the recovery variable from neuron currents, satisfying the ODE
\[
\mathbf{f}(X, \bm{\theta}, t) = \begin{pmatrix}
c(V-\dfrac{V^3}{3}+R) \\
-\dfrac{1}{c}(V-a+bR)
\end{pmatrix}
\]
where $\bm{\theta} = (a, b, c)$ are the associated parameters.
As in \cite{pmlr-v89-wenk19a,dondelinger2013ode}, the true parameters are set to $a = 0.2, b = 0.2, c = 3$, and we generate the true trajectories for this model using a numerical solver with initial conditions $V=-1$, $R=1$.

\begin{figure}
\centering
\caption{Inferred trajectories by MAGI for each component of the FN system over 100 simulated datasets. The blue shaded area represents the 95\% interval.}
\includegraphics[scale=0.32]{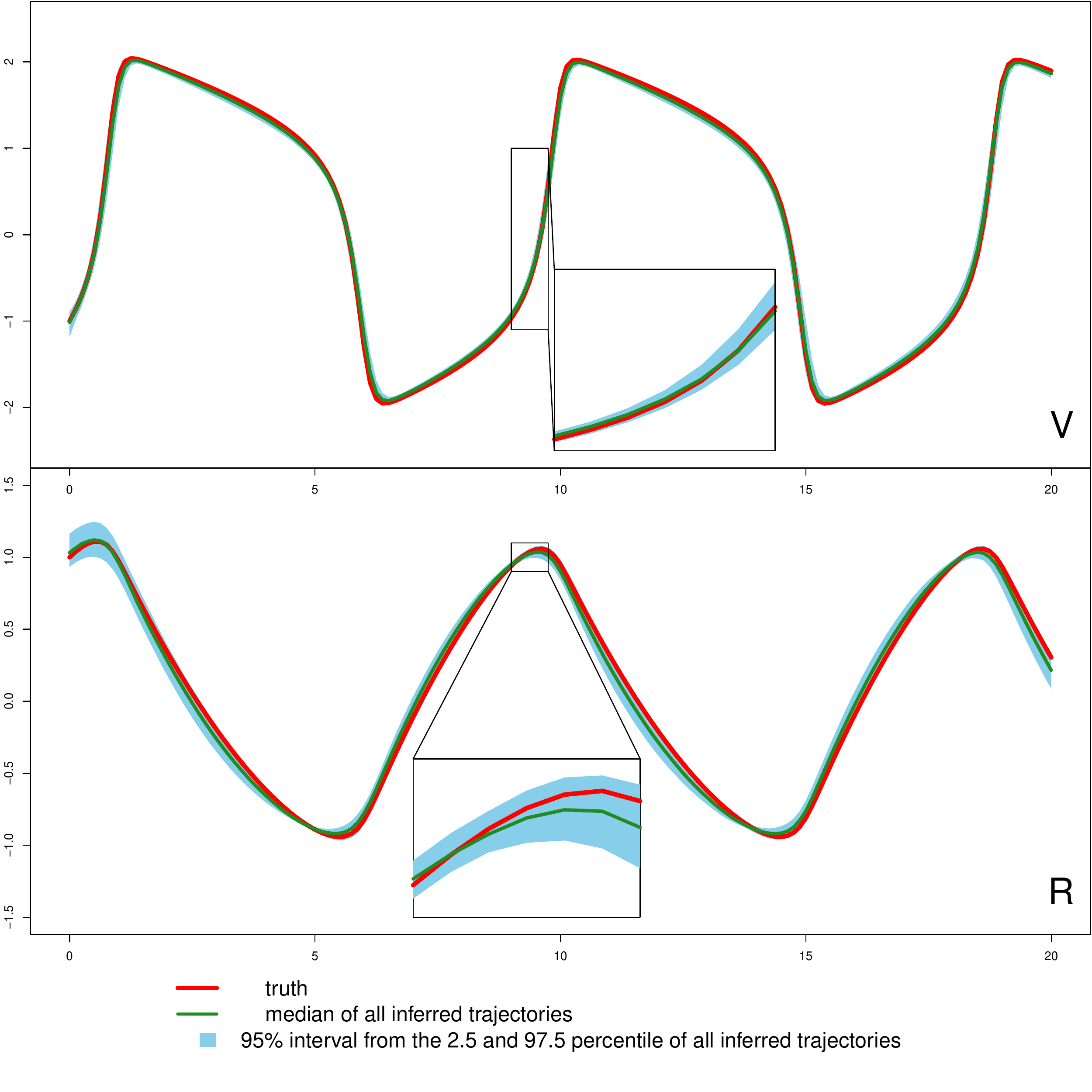}\\
\label{fig:FN}
\end{figure}

\begin{table}[ht]

\caption{Parameter inference in the FN model based on 100 simulated datasets. For each method, average parameter estimates are reported together with standard deviation; parameter RMSEs across simulations are also reported. The boldface highlights the best method in terms of parameter RMSE for each parameter. 
}\label{tab:fnparam}

\centering

\begin{tabular}{r|ll|ll|ll}
  \hline

 & \multicolumn{2}{c|}{MAGI} & \multicolumn{2}{c|}{FGPGM \cite{pmlr-v89-wenk19a}} & \multicolumn{2}{c}{AGM \cite{dondelinger2013ode} } \\
$\bm{\theta}$ & Estimate & RMSE & Estimate & RMSE & Estimate & RMSE \\
  \hline
a & 0.19 $\pm$ 0.02 & \textbf{0.02} & 0.22 $\pm$ 0.04 & 0.05 & 0.30 $\pm$ 0.03 & 0.10 \\ 
  b & 0.35 $\pm$ 0.09 & \textbf{0.17} & 0.32 $\pm$ 0.13 & 0.18 & 0.36 $\pm$ 0.06 & \textbf{0.17} \\ 
  c & 2.89 $\pm$ 0.06 & \textbf{0.13} & 2.85 $\pm$ 0.15 & 0.21 & 2.04 $\pm$ 0.14 & 0.97 \\ 
   \hline
\end{tabular}

\end{table}

To compare MAGI with FGPGM of Ref \cite{pmlr-v89-wenk19a} and AGM of Ref \cite{dondelinger2013ode}, we simulated 100 datasets under the noise setting of $\sigma_V = \sigma_R = 0.2$ with 41 observations. The noise level is chosen to be on similar magnitude with that of \cite{pmlr-v89-wenk19a}, and the noise level is set to be the same across the two components as the implementation of \cite{dondelinger2013ode} can only handle equal-variance noise. The number of repetitions (i.e., 100) is set to be the same as \cite{pmlr-v89-wenk19a} due to the high computing time of these alternative methods.

The parameter estimation results from the three methods are summarized in Table \ref{tab:fnparam}, where MAGI has the lowest parameter RMSEs among the three. Fig \ref{fig:FN} shows the inferred trajectories obtained by our method: MAGI recovers the system well, and the 95\% interval band is so narrow around the truth that we can only see the band clearly after magnification (as shown in the figure inset). 
The SI provides visual comparison of the inferred trajectories of different methods, where MAGI gives the most consistent results across the simulations. Furthermore, to assess how well the methods recover the system components, we calculated the trajectory RMSEs, and the results are summarized in Table \ref{tab:fnrmse}, where MAGI significantly outperforms the others, reducing the trajectory RMSE over the best alternative method for 60\% in $V$ and 25\% in $R$. 
We note that compared to the true parameter value, all three methods show some bias in the parameter estimates, which is partly due to the GP prior as discussed in \cite{pmlr-v89-wenk19a}, and MAGI appears to have the smallest bias.

For computing cost, the average runtime of MAGI for this system over the repetitions is 3 minutes, which is 145 times faster than FGPGM \cite{pmlr-v89-wenk19a} and 90 times faster than AGM \cite{dondelinger2013ode} on the same CPU (we follow the authors' recommendation for running their methods, see SI for details).

\begin{table}[ht]
\centering
\caption{
Trajectory RMSEs of each component in the FN system, comparing the average trajectory RMSE of the three methods over 100 simulated datasets. 
The best trajectory RMSE for each system component is shown in boldface. MAGI reduces the RMSE for 60\% in component V and 25\% in component R over the best alternative method. }\label{tab:fnrmse}

\begin{tabular}{lll}
  \hline
Method & $V$ & $R$ \\
  \hline
MAGI & \textbf{0.103} & \textbf{0.070} \\
FGPGM \cite{pmlr-v89-wenk19a} & 0.257 & 0.094 \\
AGM \cite{dondelinger2013ode}  & 1.177 & 0.662 \\
   \hline
\end{tabular}
\end{table}

\subsubsection*{Protein transduction model}
This protein transduction example is based on systems biology where components $S$ and $S_d$ represent a signaling protein and its degraded form, respectively.  In the biochemical reaction $S$ binds to protein $R$ to form the complex $S_R$, which enables the activation of $R$ into $R_{pp}$. $X = (S, S_d, R, S_R, R_{pp})$ satisfies the ODE

\[
\mathbf{f}(X, \bm{\theta}, t) = \begin{pmatrix}
-k_1 \cdot S -k_2 \cdot S \cdot R + k_3 \cdot S_R \\
k_1 \cdot S \\
-k_2 \cdot S \cdot R + k_3 \cdot S_R + \frac{V \cdot R_{pp}}{K_m + R_{pp}} \\
k_2 \cdot S \cdot R - k_3 \cdot S_R - k_4 \cdot S_R \\
k_4 \cdot S_R - \frac{V \cdot R_{pp}}{K_m + R_{pp}}
\end{pmatrix},
\]  
where $\bm{\theta} = (k_1, k_2, k_3,k_4, V, K_m)$ are the associated rate parameters.

We follow the same simulation setup as \cite{pmlr-v89-wenk19a,dondelinger2013ode}, by taking $t = \left\{0,1,2,4,5,7,10,15,20,30,40,50,60,80,100\right\}$ as the observation times, $X(0) = (1,0,1,0,0)$ as the initial values, and $\bm{\theta} = (0.07, 0.6,0.05,0.3,0.017,0.3)$ as the true parameter values.  Two scenarios for additive observation noise are considered:  $\sigma = 0.001$ (low noise) and $\sigma = 0.01$ (high noise).
Note that the observation times are unequally spaced, with only a sparse number of observations from $t=20$ to $t=100$.
Further, inference for this system has been noted to be challenging due to the non-identifiability of the parameters, in particular $K_m$ and $V$ \cite{pmlr-v89-wenk19a}.  Therefore, the parameter RMSE is not meaningful for this system, and we focus our comparison on the trajectory RMSE.

\begin{figure*}
\centering
\caption{Inferred trajectories by MAGI for each component of the protein transduction system in the high noise setting.  The red line is the truth, and the \textcolor{black}{green} line is the median inferred trajectory over 100 simulated datasets.  The blue shaded area represents the 95\% interval. The inset plots magnify the corresponding segment.}
\includegraphics[scale=0.33]{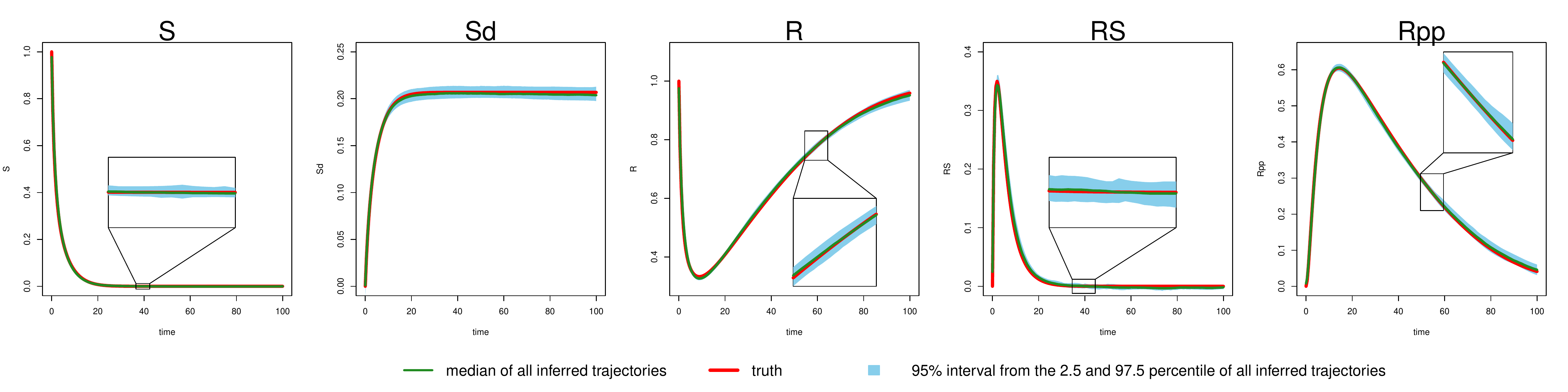}\\
\label{fig:PTtraj}
\end{figure*}

We compare MAGI with FGPGM of Ref \cite{pmlr-v89-wenk19a} and AGM of Ref \cite{dondelinger2013ode} on 100 simulated datasets for each noise setting (see the SI for method and implementation details). 
We plot the inferred trajectories of MAGI in the high noise setting in Fig \ref{fig:PTtraj}, which closely recover the system. The 95\% interval band from MAGI is quite narrow that for most of the inferred components we need magnifications (as shown in the figure insets) to clearly see the 95\% band.  We then calculated the trajectory RMSEs, and the results are summarized in Table \ref{tab:ptransrmse} for each system component. 
In both noise settings, MAGI produces trajectory RMSEs that are uniformly smaller than both FGPGM \cite{pmlr-v89-wenk19a} and AGM \cite{dondelinger2013ode} for all system components.  In the low noise setting, the advantage of MAGI is especially apparent for components $S$, $R$, $S_R$, and $R_{pp}$, with trajectory RMSEs less than half of the closest comparison method.  
For the high noise setting, MAGI reduces trajectory RMSE the most for $S_d$ and $R_{pp}$  ($\sim$50\%). 
AGM \cite{dondelinger2013ode} struggles with this example at both noise settings. 
To visually compare the trajectory RMSEs in Table \ref{tab:ptransrmse}, plots of the corresponding reconstructed trajectories by different methods at both noise settings are given in the SI. 

The runtime of MAGI for this system averaged over the repetitions is 18 minutes, which is 12 times faster than FGPGM \cite{pmlr-v89-wenk19a} and 18 times faster than AGM \cite{dondelinger2013ode} on the same CPU (we follow the authors' recommendation for running their methods, see SI for details).   

\begin{table}[ht]
\centering
\caption{
Trajectory RMSEs of the individual components in the protein transduction system, by comparing the average RMSEs of the three methods over 100 simulated datasets. The method achieving the best RMSE for each system component is shown in boldface.}

Low noise case, $\sigma=0.001$\\
\begin{tabular}{llllll}
  \hline
Method & $S$ & $S_d$ & $R$ & $S_R$ & $R_{pp}$ \\
  \hline
MAGI &  \textbf{0.0020} & \textbf{0.0013} & \textbf{0.0040 }& \textbf{0.0017} & \textbf{0.0036}\\
FGPGM \cite{pmlr-v89-wenk19a} & 0.0049 & 0.0016 & 0.0156 & 0.0036 & 0.0149 \\
AGM \cite{dondelinger2013ode} & 0.0476 & 0.2881 & 0.3992 & 0.0826 & 0.2807\\
   \hline
\end{tabular}

High noise case, $\sigma=0.01$\\
\begin{tabular}{llllll}
  \hline
Method & $S$ & $S_d$ & $R$ & $S_R$ & $R_{pp}$ \\
  \hline
MAGI & \textbf{0.0122} &  \textbf{0.0043} & \textbf{0.0167} &  \textbf{0.0135} & \textbf{0.0136}\\
FGPGM \cite{pmlr-v89-wenk19a} & 0.0128 & 0.0089 & 0.0210 &  0.0136 & 0.0309 \\
AGM  \cite{dondelinger2013ode}  & 0.0671 & 0.3125 & 0.4138 & 0.0980 & 0.2973  \\
   \hline
\end{tabular}
\label{tab:ptransrmse}
\end{table}

\section*{Discussion}

We have presented a novel methodology for the inference of dynamic systems, using manifold-constrained Gaussian processes. A key feature that distinguishes our work from the previous approaches is that it provides a principled statistical framework, firmly grounded on the Bayesian paradigm. Our method also outperformed currently available GP-based approaches in the accuracy of inference on benchmark examples. Furthermore, the computation time for our method is much faster. Our method is robust and able to handle a variety of challenging systems, including unobserved components, asynchronous observations, and parameter non-identifiability.

A robust software implementation is provided, with 
user interfaces available for R, MATLAB, and Python, as described in the SI. The user may specify custom ODE systems in any of these languages for inference with our package, by following the syntax in the examples that accompany this article. 
In practice, inference with MAGI using our software can be carried out with relatively few user interventions.  The setting of hyperparameters and initial values is fully automatic, though may be overridden by the user.  

The main setting that requires some tuning is the number of discretization points in $\bm I$.  In our examples, this was determined by gradually increasing the denseness of the points with short sampler runs, until the results become indistinguishable.  Note that further increasing the denseness of $\bm I$ has no ill effect, apart from increasing the computational time.
\textcolor{black}{To illustrate the effect of the denseness of $\bm I$ on MAGI inference results, an empirical study is included in the SI ``varying number of discretization'' section, where we examined the results of the FN model with the discretization set $\bm I$ taken to be 41, 81, 161, and 321 equally spaced points, respectively. The results confirm that our proposal of gradually increasing the denseness of $\bm I$ works well. The inference results improve as we increase $\bm I$ from 41 to 161 points, and at 161 points the results are stabilized. If we further increase the discretization to 321 points, that doubles the compute time with only a slight gain in accuracy compared to 161 points in terms of trajectory RMSEs. This empirical study 
also indicates that as $W_{\bm I}$ becomes an increasingly dense approximation of $W$, an inference limit would be expected. A theoretical study is a natural future direction of investigation.
}

\textcolor{black}{
We also investigated the stability of MAGI when 
the observation time points are farther apart. This empirical study, based on the FN model with 21 observations, is included in the SI ``FN model with fewer observations'' section. The inferred trajectories from the 21 observations are still close to the truth, while the interval bands become wider, which is expected as we have less information in this case.
We also found 
that the denseness of the discretization needs to be increased (to 321 time points in this case) to compensate for the sparser 21 observations\footnote{This finding echos the classical understanding that stiff systems require denser discretization (observations farther apart make the system appear relatively more stiff).}.
}


An inherent feature of the GP approximation is the tendency to favor smoother curves.  This limitation has been previously acknowledged \cite{pmlr-v89-wenk19a,dondelinger2013ode}.  As a consequence, two potential forms of bias can exist.  First, estimates derived from the posterior distributions of the parameters may have some statistical bias.  Second, the trajectories reconstructed by a numerical solver based on the estimated parameters may differ slightly from the inferred trajectories.  MAGI, which is built on a GP framework, does not entirely eliminate these forms of bias.  However, as seen in the benchmark systems, the magnitude of our bias in both respects is significantly smaller than the current state-of-the-art in \cite{pmlr-v89-wenk19a,dondelinger2013ode}.


We considered the inference of dynamic systems specified by ODEs in this article. \textcolor{black}{Such deterministic ODE models are often adequate to describe dynamics at the aggregate or population level \cite{kurtz1972relationship}}.
\textcolor{black}{However, when the goal is to describe the behavior of individuals (e.g., individual molecules \cite{kou2004generalized, kou2005single}), models such as stochastic differential equations (SDEs) and continuous-time Markov processes, which explicitly incorporate intrinsic (stochastic) noise, often become the model of choice. Extending our method to the inference of SDEs and continuous-time Markov models 
is a future direction we plan to investigate. Finally, recent developments in deep learning have shown connections between deep neural networks and GPs \cite{jacot2018neural,lee2018deep}.  It could thus also be interesting to explore the application of neural networks to model the ODE system outputs $\bm{x}(t)$ in conjunction with GPs.}

\section*{Materials and Methods}
For notational simplicity, we drop the dimension index $d$ in this section when the meaning is clear.


\subsection*{Algorithm overview}
We begin by summarizing the computational scheme of MAGI.  Overall, we use Hamiltonian Monte Carlo (HMC) \cite{neal2011mcmc} to obtain samples of $\bm{X_I}$ \textcolor{black}{and the parameters} from their joint posterior distribution. Details of the HMC sampling are included in the SI section `Hamiltonian Monte Carlo'. At each iteration of HMC, $\bm{X_I}$ \textcolor{black}{and the parameters\footnote{The parameters here refer to $\bm{\theta}$ and $\sigma$. If the noise level $\sigma$ is known a priori, the parameters then refer to $\bm{\theta}$ only.}} 
are updated together with a joint gradient, with leapfrog step sizes automatically tuned during the burn-in period to achieve an acceptance rate between 60-90\%. At the completion of HMC sampling (and after discarding an appropriate burn-in period for convergence), we take the posterior means of $\bm{X_I}$ as the inferred trajectories, and the posterior means of \textcolor{black}{the sampled parameters} 
as the parameter estimates. The techniques we use to temper the posterior and speed up the computations are discussed in the following `Prior tempering' subsection and `Techniques for computational efficiency' in the SI.

Several steps are taken to initialize the HMC sampler.  First, we apply a GP fitting procedure to obtain values of $\bm\phi$ and $\sigma$ for the observed components; the computed values of the hyper-parameters $\bm\phi$ are subsequently held fixed during the HMC sampling, while the computed value of $\sigma$ is used as the starting value in the HMC sampler. \textcolor{black}{(If $\sigma$ is known, the GP fitting procedure is used to obtain values of $\bm\phi$ only.)}
Second, starting values of $\bm{X_I}$ for the observed components are obtained by linearly interpolating between the observation time points. Third, starting values for the remaining quantities -- $\bm{\theta}$ and $(\bm{X_I},\bm\phi)$ for any unobserved component(s) -- are obtained by optimization of the posterior as described below.






\subsection*{Setting hyper-parameters $\bm\phi$ for observed components} 
The GP prior $ X_d(t) \sim \mathcal{GP}(\mathcal{\mu}_d, \mathcal{K}_d), ~  t \in [0, T]$, is on each component $X_d(t)$ separately.
The Gaussian process Matern kernel $\mathcal{K}(l) = \phi_1\frac{2^{1-\nu}}{\Gamma(\nu)}\left(\sqrt{2\nu}\frac{l}{\phi_2}\right)^\nu B_\nu\left(\sqrt{2\nu}\frac{l}{\phi_2}\right)$ has two hyper-parameters that are held fixed during sampling: $\phi_1$ controls overall variance level of the GP, while $\phi_2$ controls the bandwidth for how much neighboring points of the GP affect each other. 

When the observation noise level $\sigma$ is unknown, values of $(\phi_1, \phi_2, \sigma)$ are obtained jointly by maximizing GP fitting without conditioning on any ODE information, namely:
\begin{align}
(\tilde{\bm{\phi}}, \tilde{\sigma})  &= \argmax_{\bm{\phi}, \sigma} p(\bm{\phi}, \sigma^2 | \bm{y_{_{I_0}}}) \notag\\
& = \argmax_{\bm{\phi}, \sigma} \pi_{\Phi_1}(\phi_1) \pi_{\Phi_2}(\phi_2) \pi_\sigma(\sigma^2) p(\bm{y_{_{I_0}}} | \bm{\phi}, \sigma^2)
\end{align}
where $\bm{y_{_{I_0}}} |\bm{\phi}, \sigma \sim \mathcal{N}(0, \mathcal{K}_{\phi} + \sigma^2)$. The index set $I_0$ is the smallest evenly spaced set such that all observation time points in this component are in $I_0$, i.e., $\bm{\tau} \subseteq I_0$. 
The priors $\pi_{\Phi_1}(\phi_1)$ and $\pi_\sigma(\sigma^2)$ for the variance parameter $\phi_1$ and $\sigma$ are set to be flat. The prior $\pi_{\Phi_2}(\phi_2)$ for the bandwidth parameter $\phi_2$ is set to be a Gaussian distribution: \textcolor{black}{(a) the mean $\mu_{\Phi_2}$ is set to be half of the period corresponding to the frequency that is the weighted average of all the frequencies in the Fourier transform of $y$ on $I_0$ (the values of $y$ on $I_0$ are linearly interpolated from the observations at $\bm{\tau}$), where the weight on a given frequency is the squared modulus of the Fourier transform with that frequency, and (b) the standard deviation is set such that $T$ is three standard deviations away from $\mu_{\Phi_2}$. }
This Gaussian prior on $\phi_2$ serves to prevent it from being too extreme. In the subsequent sampling of $(\bm{\theta}, \bm{X_\tau}, \sigma^2)$, the hyper-parameters $\bm\phi$ are fixed at $\tilde{\bm{\phi}}$ while $\tilde{\sigma}$ gives the starting value of $\sigma$ in the HMC sampler.

If $\sigma$ is known, then values of $(\phi_1, \phi_2)$ are obtained by maximizing
\begin{equation}
\tilde{\bm{\phi}} = \argmax_{\bm{\phi}} p(\bm{\phi} | \bm{y_{_{I_0}}}, \sigma^2) = \argmax_{\bm{\phi}} \pi_{\Phi_1}(\phi_1) \pi_{\Phi_2}(\phi_2) p(\bm{y_{_{I_0}}} | \bm{\phi}, \sigma^2)
\end{equation}
and held fixed at $\tilde{\bm{\phi}}$ in the subsequent HMC sampling of $(\bm{\theta}, \bm{X_\tau})$.  The priors for $\phi_1$ and $\phi_2$ are the same as previously defined.

\subsection*{Initialization of $\bm{X_I}$ for the observed components}

To provide starting values of $\bm{X_I}$ for the HMC sampler, we use the values of $\bm{Y_\tau}$ at the observation time points and linearly interpolate the remaining points in $\bm{I}$.

\subsection*{Initialization of the parameter vector $\bm{\theta}$ when all system components are observed}
To provide starting values of $\bm{\theta}$ for the HMC sampler, we optimize the posterior \eqref{eq:real-target} as a function of $\bm{\theta}$ alone, holding $\bm{X_I}$ and $\sigma$ unchanged at their starting values, when there is no unobserved component(s). 
The optimized $\bm{\theta}$ is then used as the starting value for the HMC sampler in this case. 


\subsection*{Settings in the presence of unobserved system components: setting $\bm\phi$, initializing $\bm{X_I}$ for unobserved components, and initializing $\bm\theta$}
Separate treatment is needed for the setting of $\bm\phi$ and initialization of $(\bm\theta$, $\bm{X_I})$ for the unobserved component(s). We use an optimization procedure that seeks to maximize the full posterior in \eqref{eq:real-target} as a function of $\bm\theta$ together with $\bm\phi$ and the whole curve of $\bm{X_I}$ for unobserved components, while holding the $\sigma$, $\bm\phi$ and $\bm{X_I}$ for the observed components unchanged at their initial value discussed above. We thereby set $\bm\phi$ for the unobserved component, and the starting values of $\bm\theta$ and $\bm{X_I}$ for unobserved components at the optimized value. In the subsequent sampling, the hyper-parameters are fixed at the optimized $\bm\phi$, while the HMC sampling starts at the $\bm\theta$ and the $\bm{X_I}$ obtained by this optimization.





\subsection*{Prior tempering}
After $\bm\phi$ is set, we use a tempering scheme to control the influence of the GP prior relative to the likelihood during HMC sampling. Note that \eqref{eq:real-target} can be written as
\begin{align}
\begin{split}
& p_{{\bm{\Theta}}, \bm{X}(\bm{I})| \bm{Y(\tau)}, W_I } (\bm{\theta}, \bm{x}(\bm{I}) | \bm{y}(\bm \tau), W_{\bm{I}} = 0) \\
\propto & p_{{\bm{\Theta}}, \bm{X}(\bm{I})|  W_I } (\bm{\theta}, \bm{x}(\bm{I}) | W_{\bm{I}} = 0) p_{\bm{Y(\tau)} | \bm{X}(\bm{\tau})} ( \bm{y}(\bm \tau) | \bm{x}(\bm{\tau})).
\end{split}
\end{align}
As the cardinality of $|\bm{I}|$ increases with more discretization points, the prior part $p_{{\bm{\Theta}}, \bm{X}(\bm{I})|  W_I } (\bm{\theta}, \bm{x}(\bm{I}) | W_{\bm{I}} = 0)$ grows, while the likelihood part $p_{\bm{Y(\tau)} | \bm{X}(\bm{\tau})} ( \bm{y}(\bm \tau) | \bm{x}(\bm{\tau}))$ stays unchanged. Thus, to balance the influence of the prior, we introduce a tempering hyper-parameter $\beta$ with the corresponding posterior

\begin{align}
\begin{split}
& p_{{\bm{\Theta}}, \bm{X_{I}}|W_I,\bm{Y_\tau} }^{(\beta)} (\bm{\theta}, \bm{x_{I}} | 0, \bm{y_\tau}) \\
\propto & p_{{\bm{\Theta}}, \bm{X}(\bm{I})|  W_I } (\bm{\theta}, \bm{x}(\bm{I}) | W_{\bm{I}} = 0)^{1/\beta} p_{\bm{Y(\tau)} | \bm{X}(\bm{I})} ( \bm{y}(\bm \tau) | \bm{x}(\bm{I})) \\
\propto& \pi_{\bm{\Theta}}(\bm{\theta}) \exp\Big\{-\frac{1}{2} \sum_{d=1}^D \Big[ N_d \log(2\pi \sigma_d^{2}) +  \left\|(x_d(\bm{\tau}_d) - y_d(\bm{\tau}_d))\right\|_{\sigma_d^{-2} }^2 \notag\\
&\qquad + \frac{1}{\beta}\Big( \left\|x_d(\bm{I}) - \mu_d(\bm{I})\right\|_{C_d^{-1}}^2 \\
&\qquad\qquad\quad + \left\|\mathbf{f}_{d, \bm{I}}^{\bm{x}, \bm{\theta}} - \dot{\mu}_d(\bm{I}) - m_d (x_d(\bm{I}) - \mu_d(\bm{I}))\right\|_{K_d^{-1}}^2 \Big) \Big]\Big\} \notag
\end{split}\label{eq:termperedPost}
\end{align}

A useful setting that we recommend is $\beta = {D|\bm{I}|}/N$, where $D$ is the number of system components, $|\bm{I}|$ is the number of discretization time points, and $N=\sum_{d=1}^D N_d$ is the total number of observations. This setting aims to balance the likelihood contribution from the observations with the total number of discretization points. 

\subsection*{Data availability}
All of the data used in the article are simulation data. The details, including the models to generate the simulation data, are described in {\it Results} and the SI. Our software package also includes complete replication scripts for all the data and examples.

\subsection*{Acknowledgements}
The research of S.W.K.W. is supported in part by Discovery Grant RGPIN-2019-04771 from the Natural Sciences and Engineering Research Council of Canada. 
The research of S.C.K. is supported in part by NSF Grant DMS-1810914.

\clearpage
\renewcommand{\appendixtocname}{Supporting Information}
\renewcommand\appendixname{Supporting Information}
\renewcommand\appendixpagename{Supporting Information}

\begin{appendices}
\beginsupplement
\input{supplementarymaterial}

\end{appendices}

\end{document}

%% file: supplementarymaterial.tex
This supporting information file presents techniques for efficient computation, \textcolor{black}{a description of Hamiltonian Monte Carlo}, further details and discussion for each of the dynamic system examples in the main manuscript, \textcolor{black}{and additional empirical studies on varying the number of discretization points and reducing the number of observations}.

\section*{Techniques for computational efficiency}
After setting $\bm\phi$, the matrix inverses $C_d^{-1}$, $K_d^{-1}$ can be pre-computed and held fixed in the sampling of $\bm{X}, \bm{\theta}, \sigma$ from the target posterior, Eq. (5) in the main text. Thus, the computation of Eq. (5) in the main text at sampled values of ($\bm{X}, \bm{\theta}, \sigma$) only involves matrix multiplication, which has typical computation complexity of $O(n^2)$, where $n$ is the matrix dimension (i.e., number of discretization points). 
Due to the short-term memory and local structure of Gaussian processes (GPs), the partial correlation of two distant points diminishes quickly to zero, resulting in the off-diagonal part of precision matrices $C_d^{-1}$ and $K_d^{-1}$ being close to zero. Similarly, $m_d$ is the projection matrix of the Gaussian process to its derivative process, and since derivative is a local property, the effect from a far away point is small given one's neighboring points, resulting in the off-diagonal part of projection matrix $m_d$ being close to zero as well. Therefore, an efficient band matrix approximation may be used on $C_d^{-1}$, $K_d^{-1}$, and $m_d$ to reduce computation into $O(n)$, when calculating Eq. (5) in the main text at each sampled ($\bm{X}, \bm{\theta}, \sigma$) with a fixed band size.
In our experience, a band size of 20 to 40 is sufficient, and we recommend using an evenly spaced $\bm{I}$ for best results with the band matrix approximation and thus faster computation. In our implementation, a failure in the band approximation is automatically detected by checking for divergence in the quadratic form, and a warning is outputted to the user to increase the band size.

{
\color{black}
\section*{Hamiltonian Monte Carlo}



\subsection*{Sampling procedure with HMC}
We outline the HMC procedure for sampling from a target probability distribution.  The interested reader may refer to Ref  \cite{neal2011mcmc} for more thorough introduction to HMC.

First, suppose the target distribution has density $\pi_{\text{target}}(\bm{q}) = (1/Z)\exp(-U(\bm{q}))$, where $Z$ is the normalizing constant, and $U(\bm{q})$ is the negation of the log target density. $U(\bm{q})$ has the physical interpretation of the ``potential energy'' at ``position'' $\bm{q}$. In MAGI, $\bm{q}$ is the collection of $\bm{X_I}$ and the parameters. When the noise level $\sigma$ is known \textit{a priori}, the parameters refer to $\bm{\theta}$ only; when $\sigma$ is unknown, the parameters refer to $\bm{\theta}$ and $\sigma$. 
In MAGI the function $U(\cdot)$ is the negation of the log posterior density in Eq. (5) of the main text.

Second, momentum variables, $\bm{p}$, of the same dimension as $\bm{q}$, are introduced. A ``kinetic energy'' is defined to be $K(p) = \bm{p}^\intercal \bm{p} /2$. 

Third, define the ``Hamiltonian'' to be $H(\bm{q}, \bm{p}) = U(\bm{q}) + K(\bm{p})$, and consider the joint density of $\bm{q}$ and $\bm{p}$, which is proportional to $\exp(-H(\bm{q}, \bm{p}))$. Under this construction, $\bm{q}$ and $\bm{p}$ are independent, where the marginal probability density of $\bm{q}$ is the target $\pi_{\text{target}}$, and the marginal probability density 
of $\bm{p}$ is Gaussian. We will then sample from this augmented distribution for $(\bm{q}, \bm{p})$.

We repeat the following three steps, that together compose one HMC iteration: (1) Sample $\bm{p}$ from the normal distribution $\mathcal{N}(0, I)$ since $K(\bm{p}) = \bm{p}^\intercal \bm{p} /2$ corresponds to a Gaussian kernel; 
(2) construct a proposal $(\bm{q}^*, \bm{p}^*)$ for $(\bm{q}, \bm{p})$ by simulating the \emph{Hamiltonian dynamics} using the leapfrog method (detailed in the next subsection), and (3) accept or reject $(\bm{q}^*, \bm{p}^*)$ as the next state of $(\bm{q}, \bm{p})$ according to the usual Metropolis acceptance probability, $\min[1, \exp(-H(\bm{q}^*, \bm{p}^*) + H(\bm{q}, \bm{p}))]$.

After repeating the HMC iteration for the desired number of iterations, the sampled $\bm{q}$ are taken to be the samples from $\pi_{\text{target}}$. 
Recall $\bm{q}$ is the collection of $\bm{X_I}$ and the parameters in MAGI, so at the completion of HMC sampling, we have samples of $\bm{X_I}$ and the parameters. We finally take the posterior mean of $\bm{X_I}$ as the inferred trajectory, and the posterior means of the sampled parameters as the parameter estimates. 

\subsection*{Leapfrog method for Hamiltonian dynamics} The generating of proposals in HMC is inspired by \emph{Hamiltonian dynamics}. The leapfrog method is used to approximate the Hamiltonian dynamics. 

One step of the leapfrog method with step size $\epsilon$ from an initial point $(\bm{q}_0, \bm{p}_0)$ consists of three parts. First, we make a half step for the momentum, $\tilde {\bm{p}} = \bm{p}_0 - (\epsilon/2) \nabla U(\bm{q})|_{\bm{q} = \bm{q}_0}$.
Second, we make a full step for the position, $\bm{q}^* = \bm{q}_0 + \epsilon \tilde {\bm{p}}$.  Third, we make a full step for the momentum using the gradient evaluated at the new position, $\bm{p}^* = \tilde {\bm{p}} - (\epsilon/2) \nabla U(\bm{q})|_{\bm{q} =\bm{q}^*}$.

The step size $\epsilon$ and the number of leapfrog steps can be tuned. In our MAGI implementation, we recommend fixing the number of leapfrog steps, and tuning the leapfrog step size automatically during the burn-in period to achieve an acceptance rate between 60\% and 90\%.  

}

\section*{More details of the examples}

\subsection*{Hes1 model}


As stated in the main text, this system has three components, $X = (P,M,H)$, following the ODE
\[
\mathbf{f}(X, \bm{\theta}, t) = \begin{pmatrix}
-aPH + bM - cP \\
-dM + \frac{e}{1 + P^2} \\
-aPH + \frac{f}{1+ P^2} - gH
\end{pmatrix}
\]
where $\bm{\theta} = (a, b, c, d, e, f, g)$ are the associated parameters. 

The true parameter values in the simulation are set as $a = 0.022$, $b = 0.3$, $c = 0.031$, $d = 0.028$; $e = 0.5$, $f = 20$,  $g = 0.3$, which leads to one oscillation cycle approximately every 2 hours. The initial condition is set to be $P(0) = 1.438575$, $M(0) = 2.037488$, $H(0) = 17.90385$. Recall that these settings, along with the simulated noise level, are derived from Ref \cite{hirata2002oscillatory}, where the standard error based on repeated measures are reported to be around 15\% of the $P$ (protein) level and $M$ (mRNA) level.  Thus the simulation noise is set to be multiplicative following a log-normal distribution with standard deviation 0.15, since all components in the system are strictly positive. 
The number of observations is also set based on Ref \cite{hirata2002oscillatory}, where $P$ and $M$ are observed at 15-minute intervals for 4 hours but the $H$ component is entirely unobserved. In addition, the observations for $P$ and $M$  are asynchronous:  starting at time 0, every 15 minutes we observe $P$; starting at the 7.5 minutes, every 15 minutes we observe $M$. Following our notation in the main text, $\bm{\tau}_1 = \{0, 15, 30, \ldots, 240\}$, $\bm{\tau}_2 = \{7.5, 22.5, 37.5, \ldots, 232.5\}$, and $\bm{\tau}_3 = \varnothing$. In total we have $N_1 = 17$ observations for $P$, $N_2 = 16$ observations for $M$, and $N_3 = 0$ observations for $H$; $P$ and $M$ are never observed at the same time.  See Fig 1 (leftmost panel) of the main text for a visual illustration. 

We provide additional details on how to set up MAGI, as applied to this system.  Since the components are strictly positive, we first apply a log-transformation to the system so that the resulting noise is additive Gaussian. 
Define
\[
\tilde{P} = \log P,\quad \tilde{M} = \log M,\quad \tilde{H} = \log H,
\]
so that the transformed system is:
\[
\frac{d \tilde{\bm{X}}(t)}{dt} =  \begin{pmatrix}
-a\exp(\tilde{H}) + b\exp(\tilde{M} - \tilde{P}) - c \\
-d + e\exp(-\tilde{M})(1 + \exp(2\tilde{P}))^{-1} \\
-a\exp(\tilde{P}) + f\exp(-\tilde{H})(1 + \exp(2\tilde{P}))^{-1} - g
\end{pmatrix}.
\]
We conduct all the inference on the log-transformed system, and transform back to the original scale only at the final step to obtain inferred trajectories on the original scale.

As described in ``Setting hyper-parameters $\bm\phi$ for observed components'' in the Materials and Methods, we consider the observed $P$ component and the observed $M$ component separately when setting their respective hyper-parameters $\bm\phi$. For $P$, since the observation time points are already equally spaced, we have $I_0 = {\bm\tau}_1 = \left\{0, 15, 30, \ldots, 240\right\}$; $\tilde{\bm{\phi}}$ is obtained by optimization of Eq (8) in the main text given $\bm{y_{1,{I_0}}} = \bm{y_{1, {\tau_1}}}$, and fixing the noise level $\sigma$ at the true value of 0.15. For $M$, since the observation time points are also equally spaced, we have $I_0 = {\bm\tau}_2 = \left\{7.5, 22.5, 37.5, \ldots, 232.5\right\}$; $\tilde{\bm{\phi}}$ for $M$ is obtained by optimization of Eq (8) in the main text, given $\bm{y_{2,{I_0}}} = \bm{y_{2, {\tau_2}}}$, and fixing the noise level $\sigma$ at the true value of 0.15 as well.

Next, we consider the discretization set $\bm{I}$. In this example we use all observation time points as the discretization set, i.e., $I=\bm{\tau}_1 \cup \bm{\tau}_2  = \{0, 7.5, 15, 22.5, \ldots, 232.5, 240\}$. To initialize  $\bm{X_I}$ for the observed component $P$ and $M$, we follow the approach as described in Materials and Methods, using the values of $\bm{y_\tau}$ at the observation time points and linear interpolation for the remaining points in $\bm{I}$.


We set the hyper-parameter $\bm{\phi}$ and the initial values for the unobserved component $H$ by maximizing the full likelihood function, Eq. (5) of the main text, as described in the \textit{Materials and Methods} Section (``Settings in the presence of unobserved system components: setting $\bm\phi$, initializing $\bm{X_I}$ for unobserved components, and initializing $\bm\theta$'').

To balance the contribution from the GP prior and that from the observed data, we use prior tempering (as described in the ``Prior tempering'' subsection of \textit{Materials of Methods} of the main text). We set $\beta = {D|I|}/\sum_{d=1}^D N_d= 3$, since we have a total of 33 observations (17 observations for $P$, 16 observations for $M$, and 0 observations for $H$) and total of 33 discretization points (at times 0, 7.5, 15, ..., 240) for each of the 3 dimensions. Finally, priors for each parameter in  $\bm{\theta}$ are set to be flat on the interval $(0, \infty)$.

Having initialized the sampler for this system, we next provide details on HMC sampling to obtain our estimates of the trajectory and parameters. A total of 20000 HMC iterations were run, with the first 10000 discarded as burn-in.  Each HMC iteration uses 500 leapfrog steps, 
where the leapfrog step size is drawn randomly from a uniform distribution on $[L, 2L]$ for each iteration.  During the burn-in period, $L$ is adaptively tuned:  at each HMC iteration $L$ is multiplied by 1.005 if the acceptance rate in the previous 100 iterations is above 90\%, and $L$ is multiplied by 0.995 if the acceptance rate in the previous 100 iterations is below 60\%.
To speed up computations, we use a band matrix approximation (see `Techniques for computational efficiency' in this SI document) with band size 20.  Using the draws from the 10000 HMC iterations after burn-in, the posterior mean of $X = (P, M, H)$ is our inferred trajectory for the system components at time points in $\bm{I}$, which are generated by MAGI without using any numerical solver; the posterior mean of $\bm{\theta} = (a, b, c, d, e, f, g)$ provides our parameter estimates. 

We make comparisons with the B-spline-based penalization method of Ref \cite{ramsay2007parameter}, which provides the estimated parameters for a given dataset and ODE, but does not provide estimates for the system components (i.e., the trajectories) of the ODE. Thus, to infer the trajectories of system components implied by the method of Ref \cite{ramsay2007parameter}, we run the numerical solver for each parameter estimate (and initial values) produced by the method of Ref \cite{ramsay2007parameter} to obtain the inferred trajectories for the system components. The method of Ref \cite{ramsay2007parameter} also has hyper-parameters, in particular, the spline basis functions. The authors' R package \texttt{CollocInfer} does not provide the capability to fit spline basis functions if there are unobserved system components. Thus, to obtain results with unobserved components, we fit these spline basis functions using the true value of all system components at the observation time points in this study, which in fact gives the method of Ref \cite{ramsay2007parameter} an additional advantage than in practice: in the analysis of real data, the true value of the system components is certainly unavailable. Specifically, we used the routines in the R package \texttt{CollocInfer} by Ref \cite{ramsay2007parameter} twice: the first time, we supply the package with the fully-observed noiseless true values of all system components at the observation time points, and thus obtain the estimated B-spline basis functions as part of the package output; the second time, we supply the package with noisy data, together with the B-spline basis functions we obtained in the first run for the unobserved component, to get the final inference results. All other settings are kept at the default values in the package.

\begin{figure}[ht]
\centering
\caption{Inference for Hes1 partially observed asynchronized system on 2000 simulated datasets, comparing MAGI to the method of Ref \cite{ramsay2007parameter}. The green line is the median of the inferred trajectories across the 2000 simulated datasets. The blue shaded area represents the 95\% interval represented by the 2.5 and 97.5 percentiles of the inferred trajectories. The upper panel is the result from MAGI, and the lower panel is result from the method of Ref \cite{ramsay2007parameter}. 
} \label{fig:hes1-ramsay}
\includegraphics[scale=0.33, page=1]{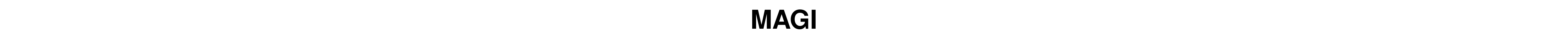}
\includegraphics[scale=0.33, trim={5in 1in 0 0.3in},clip]{charts/posteriorExpxHes1OursNoNumSolver.pdf}
\includegraphics[scale=0.33, page=4]{charts/header.pdf}
\includegraphics[scale=0.33, trim={5in 1in 0 0.3in},clip]{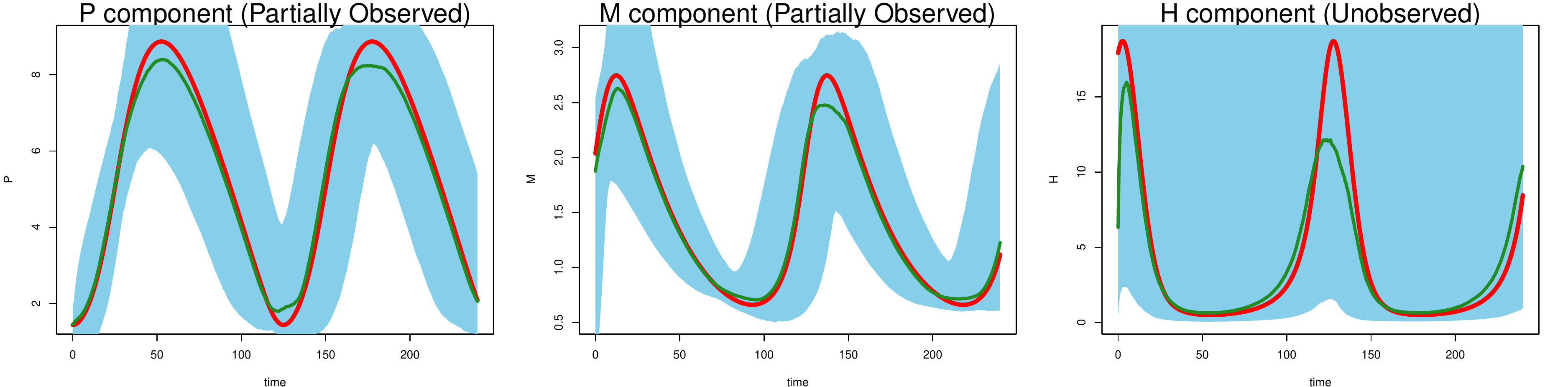}
\includegraphics[scale=0.33]{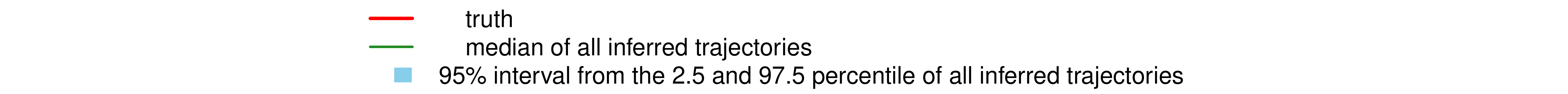}
\end{figure}

Even under this setting, the method of Ref \cite{ramsay2007parameter} had difficulty recovering the system trajectories and parameters $\bm{\theta}$ (Figure \ref{fig:hes1-ramsay}, Table 1 of the main text). Figure \ref{fig:hes1-ramsay} plots the inferred trajectories across the 2000 datasets, comparing the two methods side by side, where the method of Ref \cite{ramsay2007parameter} is seen to have difficulty to recover the unobserved component $H$. Table 1 of the main text shows the parameter RMSE, where the method of Ref \cite{ramsay2007parameter} has difficulty to recover the parameters $f$ and $g$, which are associated with the unobserved component $H$. Even for the observed components $P$ and $M$, the inferred trajectory of Ref \cite{ramsay2007parameter} has much larger RMSE compared to MAGI (see Figure \ref{fig:hes1-ramsay} and Table 2 of the main text).

Finally, we want to highlight that none of the other benchmark methods, for example, \cite{dondelinger2013ode,pmlr-v89-wenk19a}, provides software that is equipped to handle an unobserved component.

\subsection*{FitzHugh-Nagumo (FN) Model}




As stated in the main text, the FitzHugh-Nagumo (FN) model has two components, $X = (V, R)$, following the ODE
\[
\mathbf{f}(X, \bm{\theta}, t) = \begin{pmatrix}
c(V-\dfrac{V^3}{3}+R) \\
-\dfrac{1}{c}(V-a+bR)
\end{pmatrix}
\]  
where $\bm{\theta} = (a, b, c)$ are the associated parameters.

Following the same simulation setup as Refs \cite{pmlr-v89-wenk19a,dondelinger2013ode}, the initial conditions of the system are set at $X(0) = (V(0), R(0)) = (-1,1)$, the true parameter values are set at $\bm{\theta} = (a, b, c) = (0.2, 0.2, 3)$, and the system is observed at the equally spaced time points from 0 to 20 with 0.5 interval, i.e, $\bm\tau = \left\{0,0.5,1,1.5,\ldots,20\right\}$. Simulated observations have Gaussian additive noise with $\sigma = 0.2$ on both components.
 
We provide additional details on how to set up MAGI, as applied to this system. As described in ``Setting hyper-parameters $\bm\phi$ for observed components'' in the Materials and Methods, the smallest index set that includes the observation time points is $\bm{I_0} = \bm\tau = \left\{0,0.5,1,1.5,\ldots,20\right\}$; then given $ \bm{y_{_{\tau}}}$, values of $(\tilde{\bm{\phi}}, \tilde{\sigma})$ are obtained by optimizing Eq (7) in the main text. Next, we consider the discretization set $\bm{I}$. In this example we insert 3 additional equally spaced discretization time points between two adjacent observation time points, i.e., $\bm{I}=\{0,0.125,0.25 \ldots, 19.875, 20\}$, $|\bm{I}| = 161$ time points.  
As noted in the Discussion section of the main text, we successively increased the denseness of points in $\bm{I}$ and found that a further increase in the number of discretization points yielded \textcolor{black}{only slightly better} results as $\bm{I}=\{0,0.125,0.25 \ldots, 19.875, 20\}$.
Next, to initialize  $\bm{X_I}$ for the sampler, we follow the approach as described in Materials and Methods, using the values of $\bm{y_\tau}$ at the observation time points and linear interpolation for the remaining points in $\bm{I}$. Then, we obtain a starting value of $\bm{\theta}$ for the HMC sampler according to the ``Initialization of the parameter vector $\bm{\theta}$ when all system components are observed'' subsection in the main text.  We apply tempering to the posterior distribution following our guideline in the ``Prior tempering'' subsection in the main text, where $\beta = {D|\bm{I}|}/\sum_{d=1}^{D}N_d = (161\times 2)/(41\times 2)$.  Finally, the prior distributions for each parameter in  $\bm{\theta}$ are set to be flat on $(0, \infty)$.

Having initialized the sampler for this system, we run HMC sampling to obtain our estimates of the trajectory and parameters. A total of 20000 HMC iterations were run, with the first 10000 discarded as burn-in.  Each HMC iteration uses 100 leapfrog steps, 
where the leapfrog step size is drawn randomly from a uniform distribution on $[L, 2L]$ for each iteration.  During the burn-in period, $L$ is adaptively tuned:  at each HMC iteration $L$ is multiplied by 1.005 if the acceptance rate in the previous 100 iterations is above 90\%, and $L$ is multiplied by 0.995 if the acceptance rate in the previous 100 iterations is below 60\%.  To speed up computations, we use a band matrix approximation (see `Techniques for computational efficiency' in this SI document) with band size 20.  Using the draws from the 10000 HMC iterations after burn-in, the posterior mean of $X = (V, R)$ is our inferred trajectory for the system components at time points in $\bm{I}$, which are generated by MAGI without using any numerical solver; the posterior mean of $\bm{\theta} = (a, b, c)$ provides our parameter estimates.

For the two benchmark methods, we strictly follow the authors' recommendation. Specifically, for FGPGM of Ref \cite{pmlr-v89-wenk19a}, we run their provided software with all settings as recommended by the authors: the standard deviation parameter $\gamma$  there for handling potential mismatch between GP derivatives and the system is set to $3\times 10^{-4}$, a Matern52 kernel is used, and 300000 MCMC iterations are run. We treat the first half of the iterations as burn-in, and use the posterior mean as the estimate of the parameters and initial conditions.  For AGM of Ref \cite{dondelinger2013ode}, the observation noise level is assumed to be known and fixed at their true values (as this method cannot handle unknown noise level), and 300000 MCMC iterations are run.  We treat the first half of the iterations as burn-in, and use the posterior mean of the sampled values of the parameters and initial conditions as their respective estimates. 

\begin{figure*}
\centering
\caption{Histograms of the estimated $\bm{\theta}$ of the FN system over 100 simulated datasets. Three methods are compared. Upper panel: MAGI. Middle panel:  FGPGM of Ref \cite{pmlr-v89-wenk19a}. Lower panel: AGM of Ref \cite{dondelinger2013ode}. The red line is the true parameter value. 
}\label{fig:fn-parameter-histogram}
\includegraphics[scale=0.33, page=1]{charts/header.pdf}\\
\includegraphics[scale=0.33, trim={0 0.26in 0 0.19in}, clip]{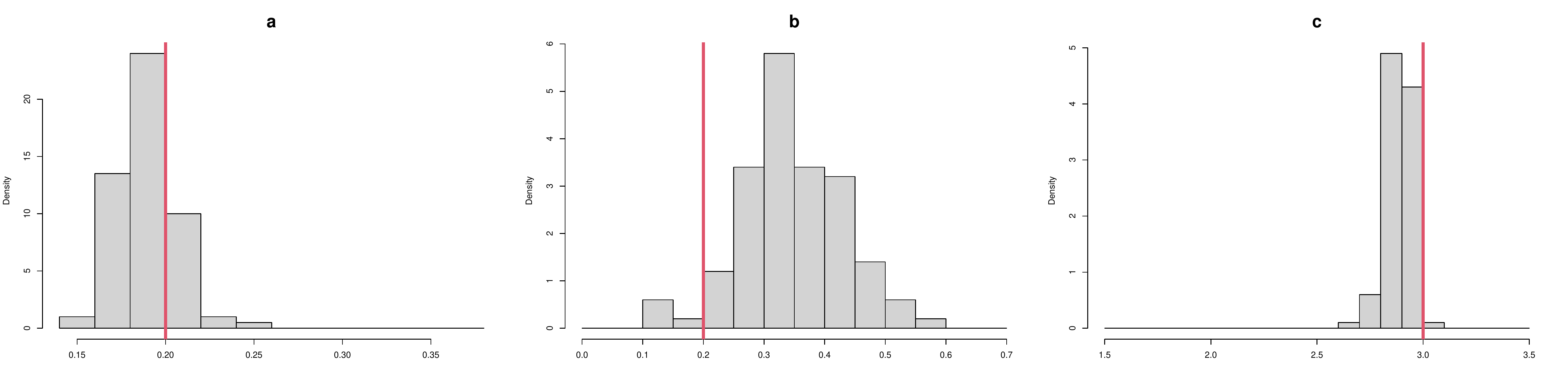}\\
\includegraphics[scale=0.33, page=2]{charts/header.pdf}\\
\includegraphics[scale=0.33, trim={0 0.26in 0 0.19in}, clip]{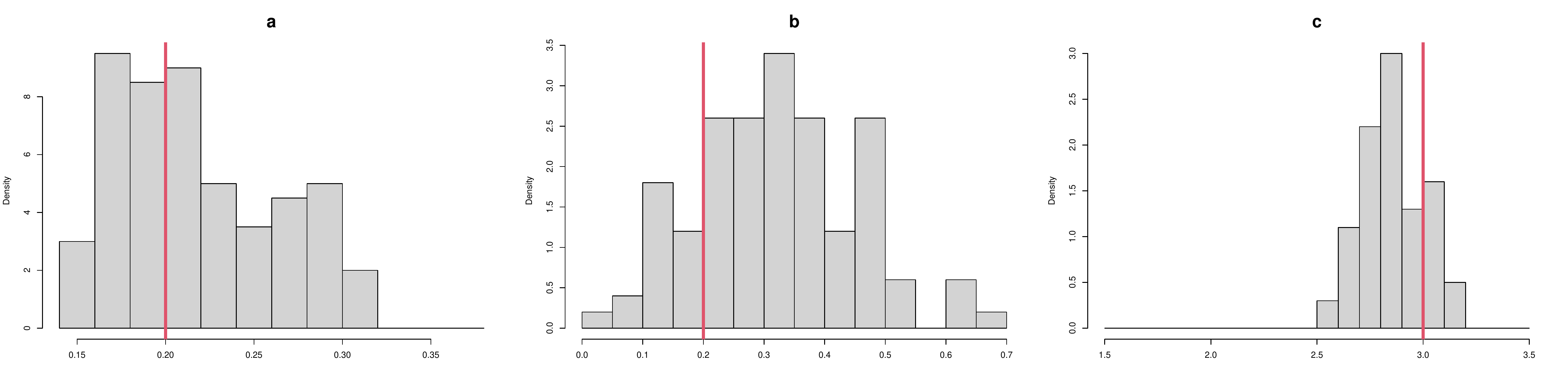}\\
\includegraphics[scale=0.33, page=3]{charts/header.pdf}\\
\includegraphics[scale=0.33, trim={0 0.26in 0 0.19in}, clip]{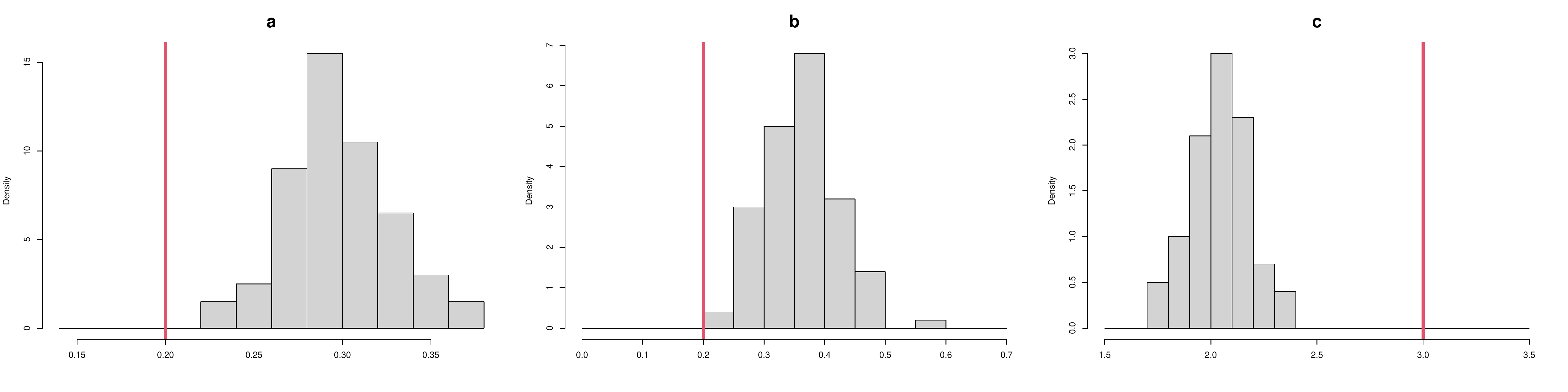}\\
\includegraphics[scale=0.33, trim={0 0.57in 0 0}, clip]{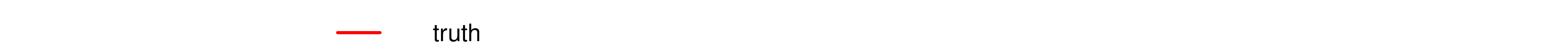}\\
\label{fig:FN-histogram}
\end{figure*}

As described in ``Metrics for assessing the quality of  system recovery'' in the main text, the \emph{parameter RMSE} is the root mean squared error (RMSE) of the parameter estimates to the true parameter value. To visualize the parameter estimates of different methods, we plot the histogram of estimated parameters for each of the methods in Figure \ref{fig:fn-parameter-histogram}.  The red line indicates the true value of each parameter $(a,b,c)$, and the histograms show the distributions of the corresponding parameter estimates over the 100 simulated datasets.  For MAGI (upper panel), the red lines lie close to the histogram values for each parameter, indicating that statistical bias is small; the spreads of the histogram values illustrate the variances of the estimates. For FGPGM \cite{pmlr-v89-wenk19a} (middle panel), the red lines lie close to the histogram values for each parameter, indicating that statistical bias is small; the spreads of the histogram values are visibly wider compared to the upper panel, showing larger variances of the estimates.  For AGM \cite{dondelinger2013ode} (lower panel), the relatively narrow spreads of the histogram values indicate that the variances of the parameter estimates are small; however, for parameters $a$ and $c$ the histogram values are much further from the true values, indicating a larger statistical bias than the other two methods.

\begin{figure*}
\centering
\caption{Reconstructed trajectories by the numerical solver for each component of the FN system from three methods. Upper panel:  MAGI. Middle panel: FGPGM of Ref \cite{pmlr-v89-wenk19a}. Lower panel: AGM of Ref \cite{dondelinger2013ode}. The red line is the true trajectory. 
The grey area is a 95\% interval represented by the 2.5 and 97.5 percentiles. }\label{fig:fn-reconstruct}
\includegraphics[scale=0.33, page=1]{charts/header.pdf}\\
\includegraphics[scale=0.33, trim={0 0.2in 0 0.5in}, clip]{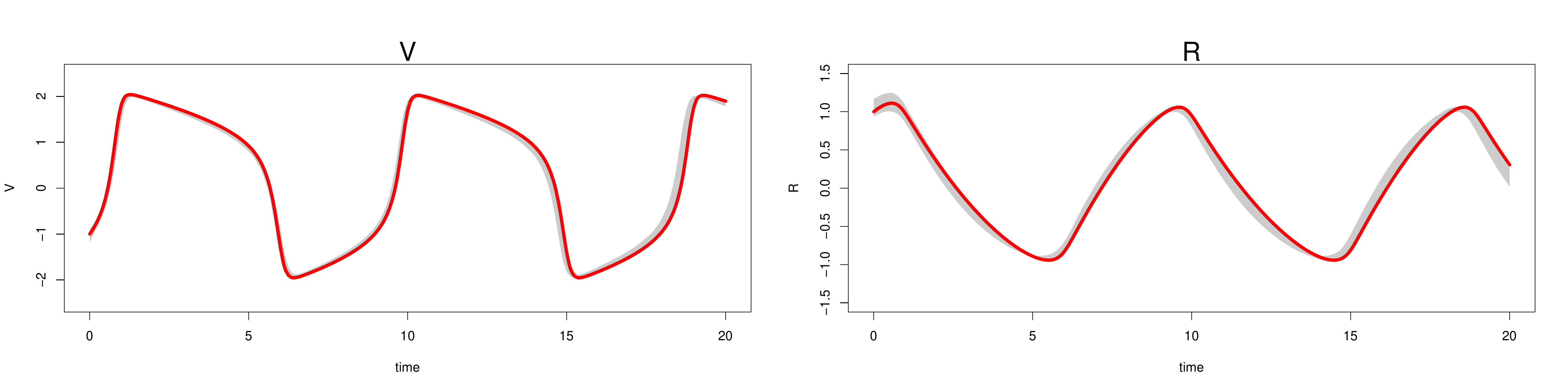}\\
\includegraphics[scale=0.33, page=2]{charts/header.pdf}\\
\includegraphics[scale=0.33, trim={0 0.2in 0 0.5in}, clip]{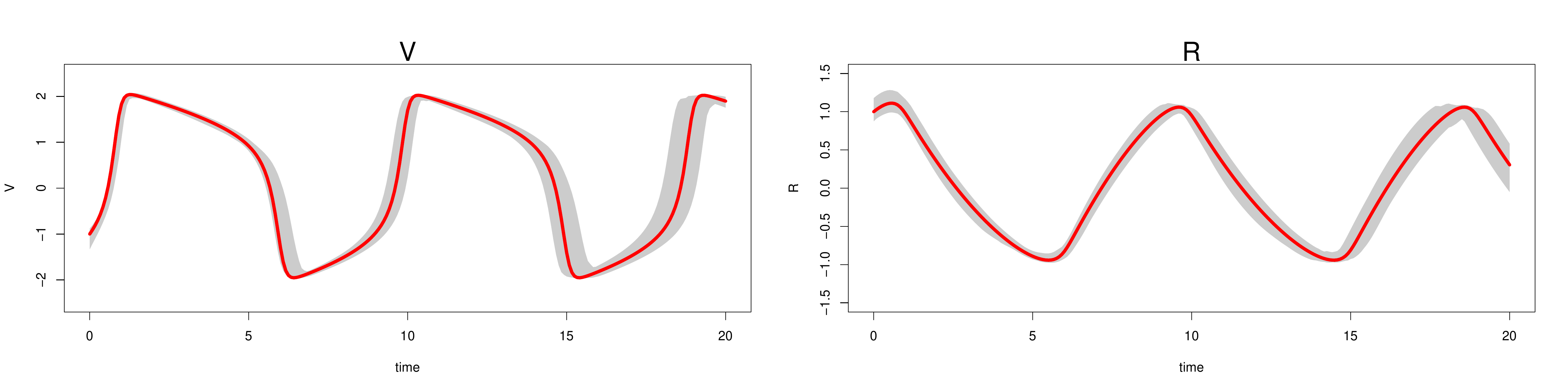}\\
\includegraphics[scale=0.33, page=3]{charts/header.pdf}\\
\includegraphics[scale=0.33, trim={0 0.2in 0 0.5in}, clip]{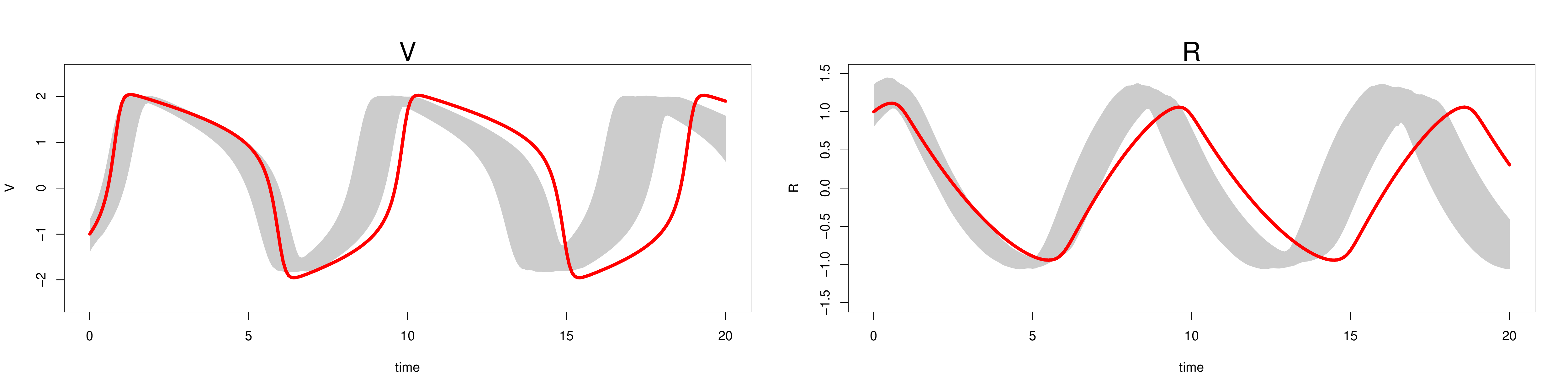}\\
\includegraphics[scale=0.33]{charts/legendGrey.pdf}\\
\label{fig:FN-reconstruct}
\end{figure*}

As described in ``Metrics for assessing the quality of system recovery'' in the main text, the \emph{trajectory RMSE} is computed for each method based on its estimate of the parameters and initial conditions. Recall that the trajectory RMSE treats the numerical ODE solution based on the true parameter values as the ground truth, and is obtained as follows: first, the numerical solver is used to reconstruct the trajectory based on the estimates of the parameter and initial condition from a given method; then, the RMSE of this reconstructed trajectory to the true trajectory at the observation time points is calculated.  To visualize the trajectory RMSEs shown in Table 4 of the main text for each method, Figure \ref{fig:fn-reconstruct} plots the true trajectory (red lines) and the 95\% interval of the reconstructed trajectories (gray bands) over the 100 simulated datasets for MAGI, FGPGM of Ref \cite{pmlr-v89-wenk19a}, and AGM of Ref \cite{dondelinger2013ode}. For MAGI (upper panel), the gray bands closely follow the true trajectories for both components, showing that the statistical bias of the reconstructed trajectories is small; the bands are also quite narrow, showing that the variance in the reconstructed trajectories is low.  For FGPGM \cite{pmlr-v89-wenk19a} (middle panel), the gray bands largely follow the true trajectories for both components, showing that the statistical bias of the reconstructed trajectories is small; however, the bands are visibly wider compared to the upper panel for both components, indicating larger variances in the reconstructed trajectories. For AGM \cite{dondelinger2013ode} (lower panel), the gray bands do not capture the true trajectory for either component, which indicates there is clear statistical bias in the reconstructed trajectories, and the bands are also much wider than the other two methods indicating a higher variance; this is probably due to the underlying statistical bias in the parameter estimates as seen in the lower panel of Figure \ref{fig:fn-parameter-histogram}.

\clearpage

\subsection*{Protein transduction model}

As stated in the main text, the protein transduction model has five components, $X = (S, S_d, R, S_R, R_{pp})$, following the ODE
\[
\mathbf{f}(X, \bm{\theta}, t) = \begin{pmatrix}
-k_1 \cdot S -k_2 \cdot S \cdot R + k_3 \cdot S_R \\
k_1 \cdot S \\
-k_2 \cdot S \cdot R + k_3 \cdot S_R + \frac{V \cdot R_{pp}}{K_m + R_{pp}} \\
k_2 \cdot S \cdot R - k_3 \cdot S_R - k_4 \cdot S_R \\
k_4 \cdot S_R - \frac{V \cdot R_{pp}}{K_m + R_{pp}}
\end{pmatrix},
\]  
where $\bm{\theta} = (k_1, k_2, k_3, k_4, V, K_m)$ are the associated rate parameters.

Following the same simulation setup as in \cite{pmlr-v89-wenk19a,dondelinger2013ode}, the initial conditions of the system are $X(0) = (1,0,1,0,0)$, the true parameter values are $\bm{\theta} = (0.07, 0.6,0.05,0.3,0.017,0.3)$, and the system is observed at the time points
 $$t = \left\{0,1,2,4,5,7,10,15,20,30,40,50,60,80,100\right\}.$$  In the low noise scenario, simulated observations have Gaussian additive noise with $\sigma = 0.001$, while in the high noise scenario $\sigma = 0.01$.  As noted in the main text, inference for this system is challenging due to the non-identifiability of the parameters, so the comparison of different method focuses on the trajectory recovery rather than the parameter RMSE.
 
We provide additional details on how to set up MAGI, as applied to this system.  Recall that the observation times are unequally spaced. Thus, as described in ``Setting hyper-parameters $\bm\phi$ for observed components'' in the Materials and Methods, we take $\bm{I_0} = \{0,1,2, \ldots, 99, 100\}$, which is the smallest index set with equally spaced time points that includes the observation times, and use linear interpolation between the observations $\bm{y_\tau}$ to obtain $\bm{y_{_{I_0}}}$; given $\bm{y_{_{I_0}}}$, values of $(\tilde{\bm{\phi}}, \tilde{\sigma})$ are obtained by optimization. 
Next, we consider the discretization set $\bm{I}$. In this example we insert 1 additional equally spaced discretization time point between two adjacent time points in $\bm{I_0}$, i.e., $\bm{I} = \{0,0.5,1 \ldots,99.5, 100\}$, $|\bm{I}| = 201$ time points. As noted in the Discussion, we successively increased the denseness of points in $\bm{I}$ and found that a further increase in the number of discretization points \textcolor{black}{did not continue to offer improved results compared to}
this setting of $\bm{I}$.
Next, to initialize  $\bm{X_I}$ for the sampler, we follow the approach as described in Materials and Methods, using the values of $\bm{y_\tau}$ at the observation time points and linear interpolation for the remaining points in $\bm{I}$. 
Then, we obtain a starting value of $\bm{\theta}$ for the HMC sampler according to ``Initialization of the parameter vector $\bm{\theta}$ when all system components are observed''.  We apply tempering to the posterior following our guideline in ``Prior tempering'', so that $\beta = {D|\bm{I}|}/\sum_{d=1}^{D}N_d = (201\times 5)/(15\times 5)$.  Finally, priors for each parameter in  $\bm{\theta}$ are set to be uniform on the interval $[0,4]$ as in Ref \cite{pmlr-v89-wenk19a}.

Having initialized the sampler for this system, we run HMC sampling to obtain samples of the trajectory and parameters. A total of 20000 HMC iterations were run, with the first 10000 discarded as burn-in.  Each HMC iteration uses 100 leapfrog steps, 
where the leapfrog step size is drawn randomly from a uniform distribution on $[L, 2L]$ for each iteration.  During the burn-in period, $L$ is adaptively tuned:  at each HMC iteration $L$ is multiplied by 1.005 if the acceptance rate in the previous 100 iterations is above 90\%, and $L$ is multiplied by 0.995 if the acceptance rate in the previous 100 iterations is below 60\%. 
To speed up computations, we use a band matrix approximation (see `Techniques for computational efficiency' in this SI document) with band size 40.  Using the draws from the 10000 HMC iterations after burn-in, the posterior mean of $X = (S, S_d, R, S_R, R_{pp})$ is our inferred trajectory for the system components, which are generated by MAGI without using any numerical solver; the posterior mean of $\bm{\theta} = (k_1, k_2, k_3, k_4, V, K_m)$ provides our parameter estimates. 

We compare MAGI with FGPGM of Ref \cite{pmlr-v89-wenk19a} and AGM of Ref \cite{dondelinger2013ode} on 100 simulated datasets for each of the two noise settings.  All methods use the same priors for $\bm{\theta}$, namely uniform on $[0,4]$ as used previously in Ref \cite{pmlr-v89-wenk19a}. 
We strictly follow the authors' recommendation for running their methods.  Specifically, for FGPGM of Ref \cite{pmlr-v89-wenk19a}, we run their provided software with all settings as recommended by the authors: the standard deviation parameter $\gamma$ there for handling potential mismatch between GP derivatives and the system is set to $10^{-4}$, a sigmoid kernel is used, and 300000 MCMC iterations are run. We treat the first half of the iterations as burn-in, and use the posterior mean as the estimate of the parameters and initial conditions.  For AGM of Ref \cite{dondelinger2013ode}, the observation noise level is assumed to be known and fixed at their true values (as this method cannot handle unknown noise level), and 300000 MCMC iterations are run.  We treat the first half of the iterations as burn-in, and use the posterior mean as the estimate of the parameters and initial conditions.

\begin{figure*}[!htbp]
\centering
\caption{Reconstructed trajectories by the numerical solver for each component of the protein transduction system from three methods, in the low noise case.  Upper panel:  MAGI.  Middle panel:  FGPGM of Ref \cite{pmlr-v89-wenk19a}.  Lower panel: AGM of Ref \cite{dondelinger2013ode}.  The red line is the true trajectory.  The grey area is the 95\% interval represented by the 2.5 and 97.5 percentiles.}
\includegraphics[scale=0.33, page=1]{charts/header.pdf}\\
\includegraphics[scale=0.33, trim={0 0.2in 0 0.2in}, clip]{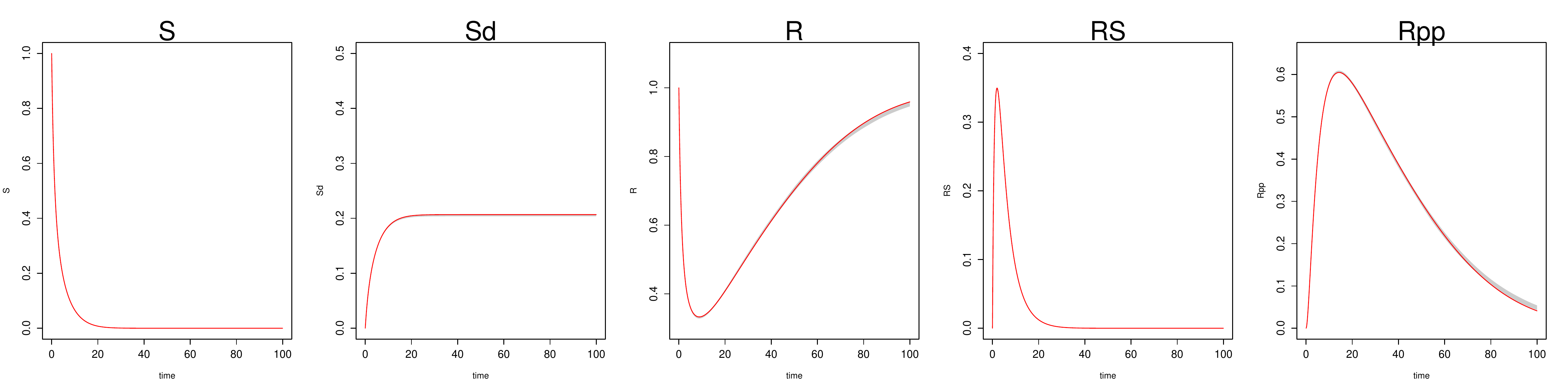}\\
\includegraphics[scale=0.33, page=2]{charts/header.pdf}\\
\includegraphics[scale=0.33, trim={0 0.2in 0 0.2in}, clip]{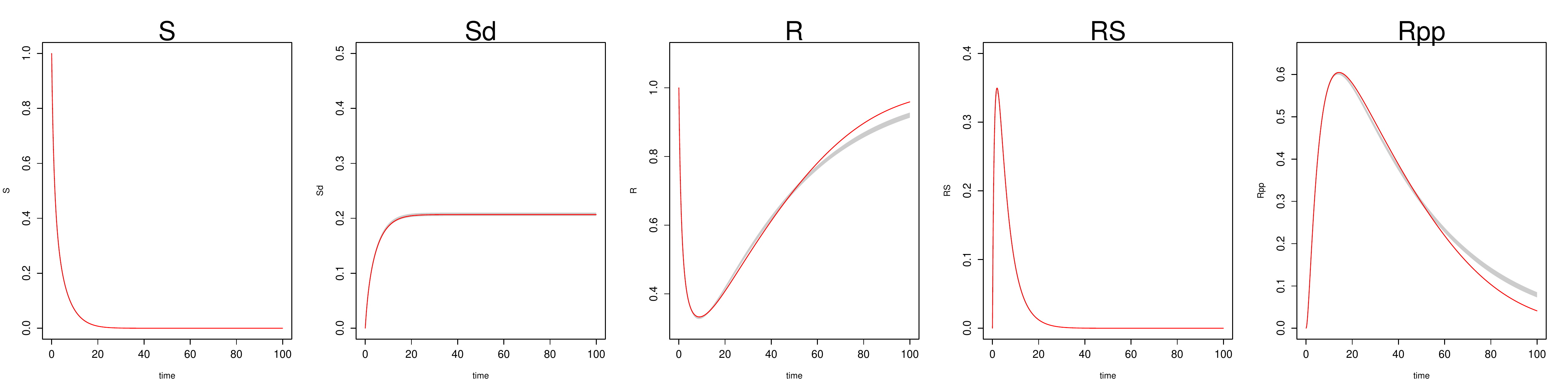}\\
\includegraphics[scale=0.33, page=3]{charts/header.pdf}\\
\includegraphics[scale=0.33, trim={0 0.2in 0 0.2in}, clip]{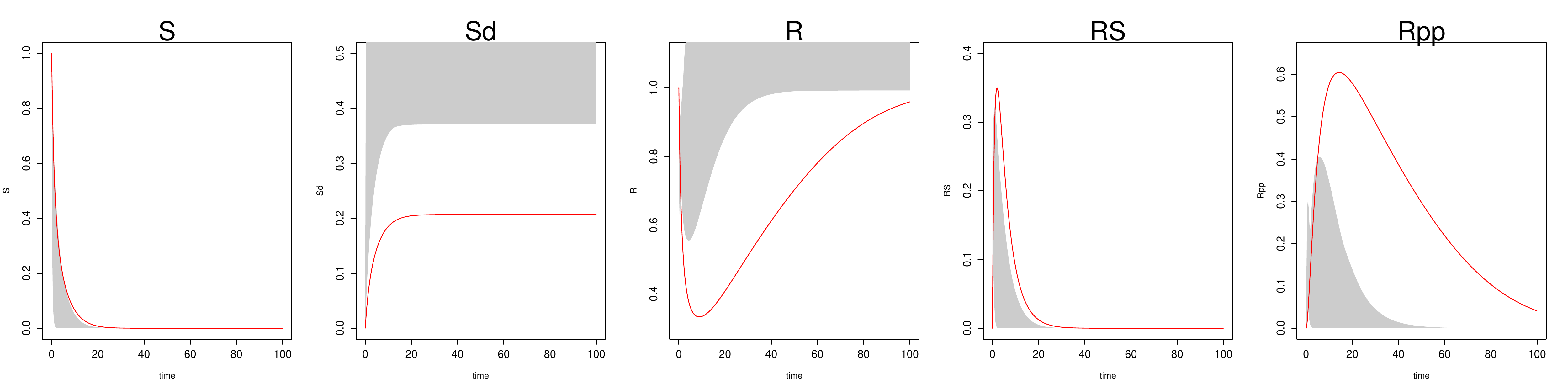}\\
\includegraphics[scale=0.33]{charts/legendGrey.pdf}\\
\label{fig:PTtrajLow}
\end{figure*}

\begin{figure*}
\centering
\caption{Reconstructed trajectories by the numerical solver for each component of the protein transduction system from three methods, in the high noise case.  Upper panel:  MAGI.  Middle panel:  FGPGM of Ref \cite{pmlr-v89-wenk19a}.  Lower panel: AGM of Ref \cite{dondelinger2013ode}.  The red line is the true trajectory.  The grey area is the 95\% interval represented by the 2.5 and 97.5 percentiles.}
\includegraphics[scale=0.33, page=1]{charts/header.pdf}\\
\includegraphics[scale=0.33, trim={0 0.2in 0 0.2in}, clip]{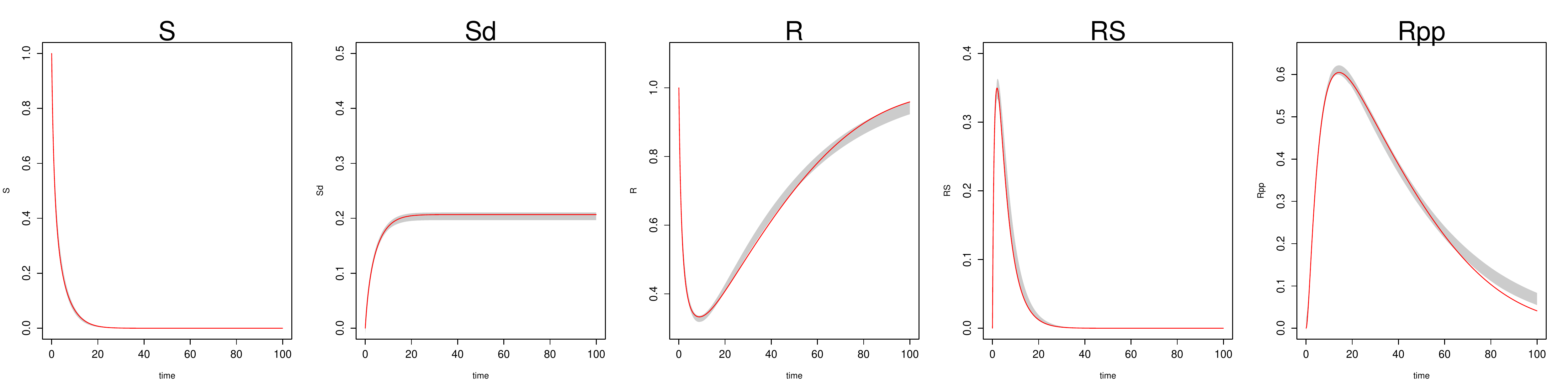}\\
\includegraphics[scale=0.33, page=2]{charts/header.pdf}\\
\includegraphics[scale=0.33, trim={0 0.2in 0 0.2in}, clip]{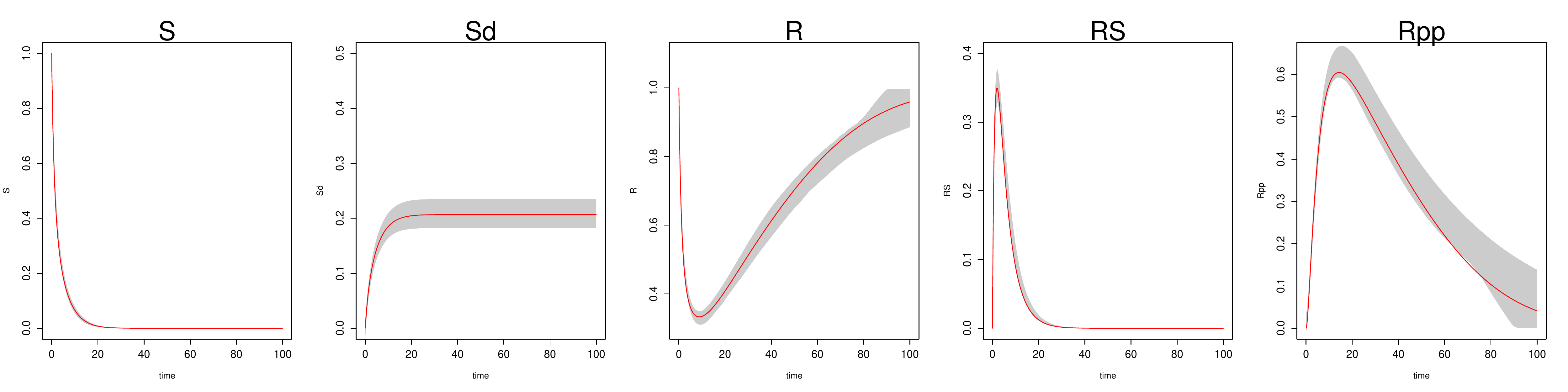}\\
\includegraphics[scale=0.33, page=3]{charts/header.pdf}\\
\includegraphics[scale=0.33, trim={0 0.2in 0 0.2in}, clip]{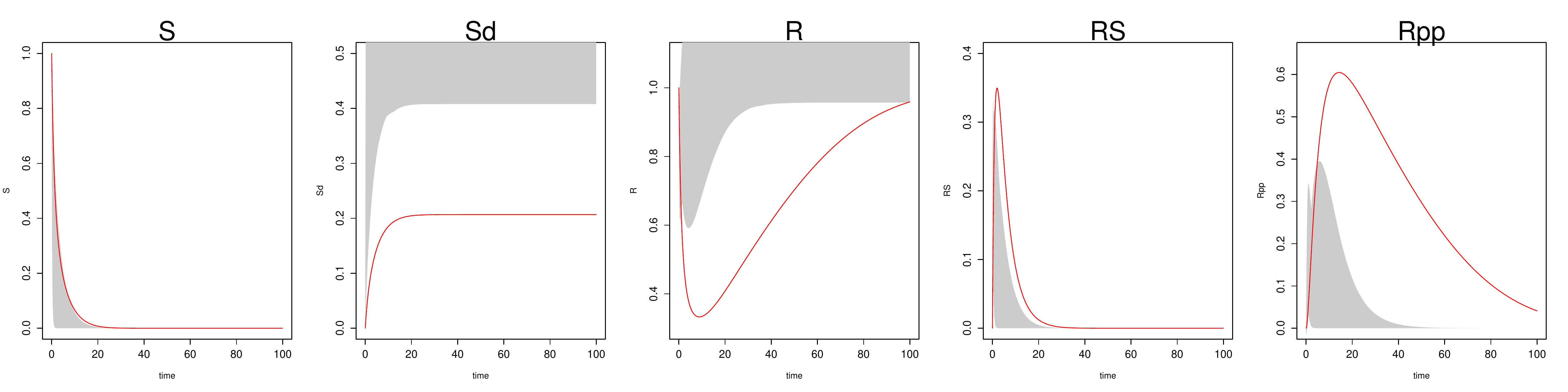}\\
\includegraphics[scale=0.33]{charts/legendGrey.pdf}\\
\label{fig:PTtrajHigh}
\end{figure*}

As described in ``Metrics for assessing the quality of  system recovery'' in the main text, the \emph{trajectory RMSE} is computed for each method based on its estimate of the parameters and initial conditions. Recall that the trajectory RMSE treats the numerical ODE solution based on the true parameter values as the ground truth, and is obtained as follows:  first, the numerical solver is used to reconstruct the trajectory based on the estimates of the parameter and initial condition from a given method; then, the RMSE of this reconstructed trajectory to the true trajectory at the observation time points is calculated. To visualize the trajectory RMSEs shown in Table 4 of the main text for each method, Figures \ref{fig:PTtrajLow} and \ref{fig:PTtrajHigh} (for the low and high noise cases, respectively) plot the true trajectory (red lines) and the 95\% interval of the reconstructed trajectories (gray bands) over the 100 simulated datasets for MAGI, FGPGM of Ref \cite{pmlr-v89-wenk19a}, and AGM of Ref \cite{dondelinger2013ode}.

In the low noise case (Figure \ref{fig:PTtrajLow}), the gray bands for MAGI (top panel) closely follow the true trajectories for all five system components, showing that the statistical bias of the reconstructed trajectories is small overall. 
The interval bands are also very narrow, indicating that the variance in the reconstructed trajectories is low.  For FGPGM \cite{pmlr-v89-wenk19a} (middle panel), the gray bands largely follow the true trajectories for most of the system components, indicating that the statistical bias of the reconstructed trajectories is small for most of the time range; however, there is clearly visible bias for the second half of the time period ($t=50$ to $t=100$) for $R$ and $R_{pp}$. The interval bands are also narrow, indicating that the variance in the reconstructed trajectories is low. For AGM \cite{dondelinger2013ode} (lower panel), the gray bands are unable to capture the true trajectories, indicating there is significant statistical bias in the reconstructed trajectories. The wide interval bands indicate a high variance in the reconstructed trajectories as well; note that the 97.5 percentile of AGM also exceeds the visible upper limit of the plots for $S_d$ and $R$.

Inference is more challenging in the high noise case (Figure \ref{fig:PTtrajHigh}).  For MAGI (upper panel), the gray bands still closely follow the true trajectories for all five system components, showing that the statistical bias of the reconstructed trajectories is small overall, with some slight bias for $R_{pp}$.  The interval bands are wider than the corresponding low noise case but still relatively narrow for all the components, indicating that the variance in the reconstructed trajectories is low.  For FGPGM \cite{pmlr-v89-wenk19a} (middle panel), the gray bands largely follow the true trajectories for all the system components, showing that the statistical bias of the reconstructed trajectories is small overall. 
The interval bands are, however, significantly wider than the upper panel; 
the variance in the reconstructed trajectories of FGPGM is thus much increased compared to that of MAGI.  For AGM \cite{dondelinger2013ode} (lower panel), the gray bands are again unable to capture the true trajectories, which indicates there is significant statistical bias in the reconstructed trajectories. The wide interval bands indicate a high variance in the reconstructed trajectories; note that the 97.5 percentile of AGM also exceeds the visible upper limit of the plots for $S_d$ and $R$.

\clearpage

{
\color{black}
\section*{Varying number of discretization}
In this section we empirically study the effect of replacing $W$ by $W_{\bm I}$. Specifically, we examine the results from varying the number of discretization points in $\bm{I}$ in the context of the FN model example.

As discussed in the main text, the number of discretization points in $\bm I$ is the main setting that requires some tuning.  In our examples, this was determined by gradually increasing the denseness of the points with short sampler runs, until the results become stabilized. 
Note that further increasing the denseness of $\bm I$ has no ill effect, apart from increasing the computational time.

Here we illustrate the effect of $\bm I$ by varying the number of discretization points, using the same dataset of the FN system  presented in the main text. The result is summarized in Table \ref{tab:fn-discretization-sensitivity}. The results in the main text Tables 3 and 4 are based on 161 discretization points. As can be seen, the inference results improve as we increase $\bm{I}$ from 41 to 161 points, and at 161 points the results are stabilized.  If we further increase the discretization to 321 points, that doubles the compute time with only a slight gain in accuracy compared to 161 points in terms of the 
RMSEs.

\begin{table}[h!]
\color{black}
    \caption{Results of FN model inference based on the same 100 simulated datasets as in the main text, with varying number of discretization points (41, 81, 161, 321) equally spaced in time. The results presented in the main text use 161 discretization points.}
    \label{tab:fn-discretization-sensitivity}
\resizebox{0.98\columnwidth}{!}{
    \centering
\begin{tabular}{c|ll|ll|ll|cc|r}
  \hline
number of & \multicolumn{2}{c|}{parameter a} & \multicolumn{2}{c|}{parameter b} & \multicolumn{2}{c|}{parameter c} & \multicolumn{2}{c|}{trajectory RMSE} & \multicolumn{1}{c}{run time}\\
discretizations & Estimate & RMSE & Estimate & RMSE & Estimate & RMSE & V & R & \multicolumn{1}{c}{(minutes)}\\
  \hline
   41 & 0.20 $\pm$ 0.03 & 0.026 & 0.24 $\pm$ 0.08 & 0.091 & 2.83 $\pm$ 0.12 & 0.211 & 0.358 & 0.146 & 0.84 \\ 
   81 & 0.19 $\pm$ 0.02 & 0.020 & 0.34 $\pm$ 0.09 & 0.165 & 2.82 $\pm$ 0.07 & 0.199 & 0.270 & 0.142 & 1.67 \\ 
  161 & 0.19 $\pm$ 0.02 & 0.020 & 0.35 $\pm$ 0.09 & 0.172 & 2.89 $\pm$ 0.06 & 0.128 & 0.103 & 0.070 & 3.13 \\ 
  321 & 0.19 $\pm$ 0.02 & 0.020 & 0.33 $\pm$ 0.09 & 0.162 & 2.92 $\pm$ 0.06 & 0.097 & 0.072 & 0.051 & 5.94 \\ 
   \hline
\end{tabular}
}
\end{table}

\clearpage

\section*{FN model with fewer observations}
In this section we study the FN system with 21 observations, which is fewer than the 41 observations presented in the main text. This investigation aims to answer two questions: (1) how does MAGI perform when the number of observations is more sparse, and (2) how does MAGI perform if the observation time points are spaced farther apart?

Following the same setup as the FN system in the main text, 
we now consider the scenario where 21 observations are made at equally spaced time points from 0 to 20, i.e, $\bm{\tau} = \left\{0,1,\ldots,20\right\}$.
When applying MAGI, the discretization set $I$ was determined by successively increasing its denseness (with short sampler runs), until the results become stabilized. The numerical results show that in this scenario with sparser observations that are also farther apart, a higher number of discretization points is needed for the results to be stabilized.  Specifically for this example with 21 observations, 321 points in the discretization set $\bm{I}$, i.e.,  $\bm{I} = \{0, 0.0625, 0.125, \ldots, 20\}$ are needed to obtain stable inference results. The required increase in discretization seen here echos the classical understanding that stiff systems require denser discretization (observations farther apart make the system appear relatively more stiff).

The inference results are presented in Table \ref{tab:fn21}. The trajectory RMSE is 0.128 for V component and 0.107 for R component, which is roughly $\sqrt{2}$ times the trajectory RMSE for that of 41 observations as reported in the main text. The $\sqrt{2}$ factor is expected, as we halved the number of observations. Further visualization in Figure \ref{fig:fn21} suggests that the inferred trajectory is quite close to the true system, while the interval bands become wider, which is expected as we have less information in this case. 
 


\begin{table}[h!]
\color{black}
    \caption{Results of FN model inference based on 100 simulated datasets, each with 21 observations. Average parameter estimates are reported together with standard deviations; parameter RMSEs across simulations are also reported; trajectory RMSEs for the two components are reported as well. The true parameters are set to $a = 0.2, b = 0.2, c = 3$, as in the main text.}
    \label{tab:fn21}
\resizebox{0.99\columnwidth}{!}{
    \centering
\begin{tabular}{c|c|ll|ll|ll|cc|r}
  \hline
number of & number of & \multicolumn{2}{c|}{parameter a} & \multicolumn{2}{c|}{parameter b} & \multicolumn{2}{c|}{parameter c} & \multicolumn{2}{c|}{trajectory RMSE} & \multicolumn{1}{c}{run time}\\
observations & discretizations & Estimate & RMSE & Estimate & RMSE & Estimate & RMSE & V & R & \multicolumn{1}{c}{(minutes)}\\
  \hline
21 & 321 & 0.19 $\pm$ 0.03 & 0.029 & 0.44 $\pm$ 0.15 & 0.280 & 2.79 $\pm$ 0.16 & 0.261 & 0.128 & 0.107 & 5.81 \\ 
   \hline
\end{tabular}
}
\end{table}

\begin{figure}[ht]
\centering
\caption{Inferred trajectories by MAGI for each component of the FN system over 100 simulated datasets, each with 21 observations. The red line is the truth, and the green line is the median inferred trajectory over 100 simulated datasets. The blue shaded area represents the 95\% interval. The black dots indicate the observations across 100 simulated datasets.} 
\label{fig:fn21}
\includegraphics[scale=0.33, page=1]{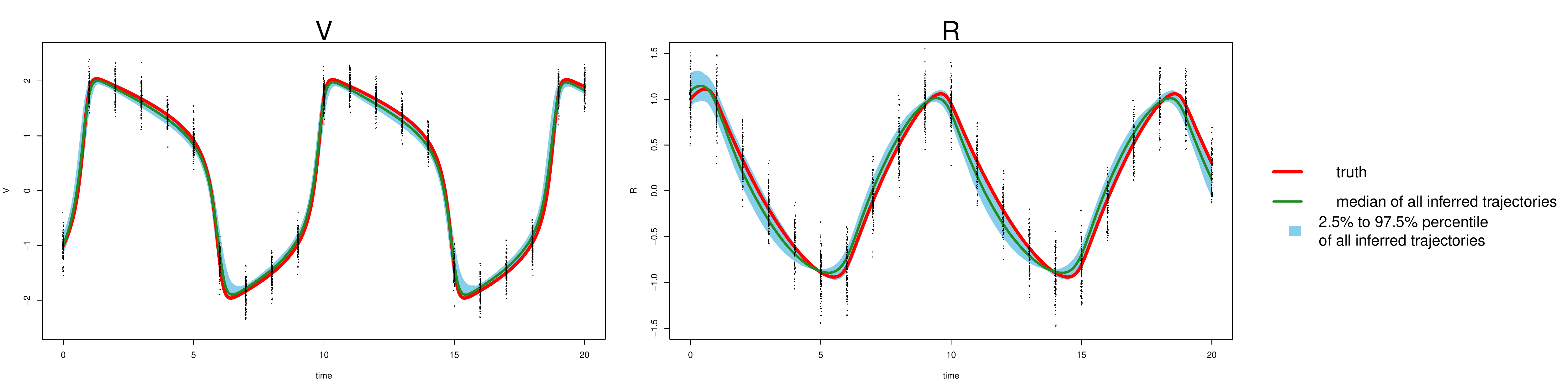}
\end{figure}

}

\clearpage

\section*{Software implementation}

User interfaces for MAGI are available for R, MATLAB, and Python at the Github repository \url{https://github.com/wongswk/magi}.  Detailed instructions are provided therein for using our package with custom ODE systems specified in any of these languages.  Detailed instructions are also provided for replicating all of our results and figures provided in the paper.